\documentclass[10pt]{article}
\usepackage{graphicx}
\usepackage{amsmath}
\usepackage{amssymb}
\usepackage{caption2}
\begin{document}
\bibliographystyle{prsty}
\begin{center}
{\large {\bf \sc{  Reanalysis of  the $Z_c(4020)$, $Z_c(4025)$, $Z(4050)$ and $Z(4250)$ as   tetraquark states
with  QCD sum rules }}} \\[2mm]
Zhi-Gang  Wang \footnote{E-mail: zgwang@aliyun.com.  }     \\
 Department of Physics, North China Electric Power University, Baoding 071003, P. R. China
\end{center}

\begin{abstract}
In this article, we  calculate the contributions of the vacuum condensates up to dimension-10  in the operator product expansion, and  study the  $C\gamma_\mu-C\gamma_\nu$ type scalar, axial-vector and tensor tetraquark states in details with the QCD sum rules. In calculations,  we use the  formula $\mu=\sqrt{M^2_{X/Y/Z}-(2{\mathbb{M}}_c)^2}$  to determine  the energy scales of the QCD spectral densities. The predictions  $M_{J=2} =\left(4.02^{+0.09}_{-0.09}\right)\,\rm{GeV}$, $M_{J=1} =\left(4.02^{+0.07}_{-0.08}\right)\,\rm{GeV}$  favor assigning  the $Z_c(4020)$ and $Z_c(4025)$   as the $J^{PC}=1^{+-}$ or $2^{++}$   diquark-antidiquark type tetraquark states, while the prediction $M_{J=0}=\left(3.85^{+0.15}_{-0.09}\right)\,\rm{GeV}$ disfavors assigning  the  $Z(4050)$ and $Z(4250)$ as the $J^{PC}=0^{++}$ diquark-antidiquark type tetraquark states. Furthermore, we discuss the strong decays of the $0^{++}$, $1^{+-}$, $2^{++}$ diquark-antidiquark type tetraquark states in details.
 \end{abstract}



\section{Introduction}

In 2008, the Belle collaboration  reported the first observation of two resonance-like structures in the $\pi^+ \chi_{c1}$ invariant mass distribution near $4.1\,\rm{ GeV}$ in the exclusive $\bar{B}^0 \to K^{-} \pi^{+} \chi_{c1}$ decays, and determined the masses and widths $M_{Z(4050)} =\left( 4051 \pm 14 {}^{+20}_{-41}\right)\,\rm{ MeV}$, $M_{Z(4250)} = \left(4248^{ +44}_{-29}{}^{ +180}_{-35}\right)\,\rm{ MeV}$, $\Gamma_{Z(4050)} = \left(82^{ +21}_{-17}{}^{ +47}_{-22}\right)\,\rm{ MeV}$  and $\Gamma_{Z(4250)} =\left(177^{ +54}_{-39}{}^{ +316}_{-61}\right)\,\rm{ MeV}$, respectively  \cite{Belle0806}. There have been several tentative assignments, such as the tetraquark states \cite{Wang4250,Wang4250-EPJC,Tetraquark4250}, molecular states \cite{Molecule4250}, etc. In 2011, the BaBar collaboration   searched  for the $Z^+(4050)$   and $Z^+(4250)$   states in the decays $\bar{B}^0 \to \chi_{c1} K^- \pi^+$   and
$ B^+ \to \chi_{ c1} K^0_S \pi^+$  based on the data collected with the BaBar detector at the SLAC PEP-II asymmetric-energy $e^+e^-$  collider, and observed no evidences \cite{BaBar1111}. The  $Z^+(4050)$   and $Z^+(4250)$   still need confirmation.

 In 2013, the BESIII collaboration  observed
the $Z^{\pm}_c(4025)$ near the $(D^{*} \bar{D}^{*})^{\pm}$ threshold in the $\pi^\mp$ recoil mass spectrum  in the process $e^+e^- \to (D^{*} \bar{D}^{*})^{\pm} \pi^\mp$ at a center-of-mass energy of $4.26\,\rm{GeV}$, and determined
 the mass and width  $M_{Z_c(4025)}=(4026.3\pm2.6\pm3.7)\,\rm{MeV}$  and $\Gamma_{Z_c(4025)}=(24.8\pm5.6\pm7.7)\,\rm{MeV}$, respectively \cite{BES1308}.
Furthermore, the  BESIII collaboration observed the  $Z_c(4020)$   in the $\pi^\pm h_c$ mass spectrum in the process $e^+e^- \to \pi^+\pi^- h_c$ at center-of-mass energies  $(3.90-4.42)\,\rm{GeV}$, and determined the  mass and width  $M_{Z_c(4020)}=(4022.9\pm 0.8\pm 2.7)\,\rm{MeV}$   and $\Gamma_{Z_c(4020)}=(7.9\pm 2.7\pm 2.6)\,\rm{MeV}$, respectively  \cite{BES1309}. There have been several tentative assignments of the  $Z_c(4025)$ and $Z_c(4020)$, such as the re-scattering effects \cite{Rescatter}, molecular states \cite{Molecule}, tetraquark states \cite{Tetraquark-Qiao}, etc.

The S-wave $D^{*} \bar{D}^{*}$ systems  have the quantum numbers $J^{PC}=0^{++}$, $1^{+-}$, $2^{++}$,  the S-wave $ \pi^\pm h_c$ systems have the quantum numbers $J^{PC}=1^{--}$, the S-wave $ \pi^\pm \chi_{c1}$ systems have the quantum numbers $J^{PC}=1^{-+}$ (or $J^{PC}=1^{--}$). On the other hand, it is also possible for the P-wave $\pi^{\pm} h_c$ and $ \pi^\pm \chi_{c1}$ systems to have the quantum numbers $J^{PC}=0^{++}$, $1^{+-}$, $2^{++}$.
In Ref.\cite{Wang1311},  we observe that the predictions based on the QCD sum rules  disfavor assigning  the $Z_c(4020)$ and $Z_c(4025)$ as the
diquark-antidiquark type vector tetraquark states.
The $Z_c(4020)$ and $Z_c(4025)$ are potential candidates of the axial-vector and tensor  tetraquark states \cite{Wang1311}.
The $Z(4050)$ and $Z(4250)$ are potential candidates of the scalar  tetraquark states \cite{Wang4250,Wang4250-EPJC}. However, we cannot exclude the possibilities that the $Z_c(4025)$, $Z_c(4020)$ and $Z(4050)$ are the same particle with the $J^{PC}=1^{+-}$ or $2^{++}$.  The $Z_c(4025)$, $Z_c(4020)$, $Z(4050)$ and $Z(4250)$ are charged charmonium-like states, their quark constituents must be $c\bar{c}u\bar{d}$ or $c\bar{c}d\bar{u}$ irrespective of the diquark-antidiquark type or meson-meson type substructures. The interested reader can consult Ref.\cite{Swanson2006} for more articles  on the exotic  $X$, $Y$ and $Z$ particles.

In Refs.\cite{Wang1311,WangHuangTao}, we distinguish
the charge conjugations of the interpolating  currents, calculate the contributions of the vacuum condensates up to dimension-10
 in the operator product expansion, study the diquark-antidiquark type axial-vector and vector tetraquark states in a systematic way  with the QCD sum rules, make possible assignments of the $X(3872)$, $Z_c(3900)$, $Z_c(3885)$, $Z_c(4020)$, $Z_c(4025)$, $Y(4360)$, $Y(4630)$ and $Y(4660)$.  Furthermore, we    explore the energy scale dependence of the hidden charmed tetraquark states  in details for the first time, and suggest a  formula
\begin{eqnarray}
\mu&=&\sqrt{M^2_{X/Y/Z}-(2{\mathbb{M}}_c)^2} \, ,
 \end{eqnarray}
 with the effective $c$-quark mass ${\mathbb{M}}_c=1.8\,\rm{GeV}$ to determine the energy scales of the  QCD spectral densities in the QCD sum rules. The  masses of the $J^{PC}=1^{-+}$,$1^{--}$  tetraquark states with symbolic quark structure $c\bar{c}u\bar{d}$  disfavor assigning  the $Z_c(4020)$, $Z_c(4025)$ and $Y(4360)$
as the vector tetraquark states; the masses of the vector  tetraquark states with symbolic quark structures $c\bar{c}s\bar{s}$ and $c\bar{c}(u\bar{u}+d\bar{d})/\sqrt{2}$  favor  assigning the $Y(4660)$ (or $Y(4630)$) as the $1^{--}$ tetraquark state \cite{Wang1311}.

In Refs.\cite{Wang4250,Wang4250-EPJC}, we study the $C\gamma_\mu-C\gamma^\mu$ and $C\gamma_5-C\gamma_5$ type scalar  tetraquark states with the QCD sum rules by  carrying  out the operator product expansion to the vacuum condensates  up to
dimension-10 and setting the energy scale to be $\mu=1\,\rm{GeV}$.
In Ref.\cite{Tetraquark-Qiao}, the Qiao and Tang study the   vector and tensor tetraquark states with the QCD sum rules by carrying   out the operator product expansion to the vacuum condensates up to dimension-8. They  try to assign the $Z_c(4025)$ as the $J^P=2^+$ tetraquark state (to be more precise, the $J^{PC}=2^{+-}$ tetraquark state),  but do not show (or do not specify)  the energy scales of the QCD spectral densities.
In Refs.\cite{Wang4250,Wang4250-EPJC,Tetraquark-Qiao},  some higher dimension vacuum condensates involving the gluon condensate, mixed condensate and four-quark condensate are neglected, which maybe impair the predictive ability.

 In this article, we distinguish the charge conjugations of the interpolating currents,  calculate the contributions of the vacuum condensates up to dimension-10  in a  consistent way,  study the  scalar, axial-vector, tensor  hidden charmed tetraquark states with the $C\gamma_\mu-C\gamma_\nu$ type interpolating currents in a systematic way, make tentative  assignments of the $Z_c(4020)$, $Z_c(4025)$, $Z(4050)$ and $Z(4250)$ based on the QCD sum rules.
 Furthermore, we explore  validity of the formula on how to determine the energy scales of the QCD spectral densities \cite{Wang1311}.

 The diquarks have  five Dirac tensor structures, scalar $C\gamma_5$,
pseudoscalar $C$, vector $C\gamma_\mu \gamma_5$, axial vector
$C\gamma_\mu $  and  tensor $C\sigma_{\mu\nu}$. The structures
$C\gamma_\mu $ and $C\sigma_{\mu\nu}$ are symmetric, the structures
$C\gamma_5$, $C$ and $C\gamma_\mu \gamma_5$ are antisymmetric. The
attractive interactions of one-gluon exchange  favor  formation of
the diquarks in  color antitriplet $\overline{3}_{ c}$, flavor
antitriplet $\overline{3}_{ f}$ and spin singlet $1_s$ (or flavor
sextet  $6_{ f}$ and spin triplet $3_s$) \cite{One-gluon}, the favored configurations are the scalar and axial-vector diquark states.
  We take the diquark states as the basic constituents   following
Jaffe and Wilczek \cite{Jaffe2003}. If additional partial derivative is not introduced,  the scalar $C\gamma_5$ diquark-antidiquark pair can form scalar tetraquark states only, while the axial-vector $C\gamma_\mu$ diquark-antidiquark pair can form scalar, axial-vector, tensor tetraquark states. The scalar and axial-vector heavy-light diquark states have almost  degenerate masses from the QCD sum rules \cite{WangDiquark}. In this article, we choose the $C\gamma_\mu-C\gamma_\nu$  type  interpolating currents to study the  tetraquark states.

The article is arranged as follows:  we derive the QCD sum rules for the masses and pole residues of  the scalar, axial-vector, tensor tetraquark states  in section 2; in section 3, we present the numerical results and discussions; section 4 is reserved for our conclusion.

\section{QCD sum rules for  the  scalar, axial-vector and tensor tetraquark states }
In the following, we write down  the two-point correlation functions $\Pi_{\mu\nu\alpha\beta}(p)$ and $\Pi(p)$ in the QCD sum rules,
\begin{eqnarray}
\Pi_{\mu\nu\alpha\beta}(p)&=&i\int d^4x e^{ip \cdot x} \langle0|T\left\{\eta_{\mu\nu}(x)\eta_{\alpha\beta}^{\dagger}(0)\right\}|0\rangle \, , \\
\Pi(p)&=&i\int d^4x e^{ip \cdot x} \langle0|T\left\{\eta(x)\eta^{\dagger}(0)\right\}|0\rangle \, ,
\end{eqnarray}

\begin{eqnarray}
 \eta_{\mu\nu}(x)&=&\frac{\epsilon^{ijk}\epsilon^{imn}}{\sqrt{2}}\left\{u^j(x)C\gamma_\mu c^k(x) \bar{d}^m(x)\gamma_\nu C \bar{c}^n(x)+tu^j(x)C\gamma_\nu c^k(x)\bar{d}^m(x)\gamma_\mu C \bar{c}^n(x) \right\} \, , \\
  \eta(x)&=&\epsilon^{ijk}\epsilon^{imn}u^j(x)C\gamma_\mu c^k(x) \bar{d}^m(x)\gamma^\mu C \bar{c}^n(x) \, ,
\end{eqnarray}
where the $i$, $j$, $k$, $m$, $n$ are color indexes, the $C$ is the charge conjugation matrix.
 Under charge conjugation transform $\widehat{C}$, the currents $\eta_{\mu\nu}(x)$ and $\eta(x)$ have the following properties,
\begin{eqnarray}
\widehat{C}\,\eta_{\mu\nu}(x)\,\widehat{C}^{-1}&=&\pm \,\eta_{\mu\nu}(x)\mid_{u\leftrightarrow d } \,\,\,\, {\rm for}\,\,\,\, t=\pm1\, , \nonumber \\
\widehat{C}\,\eta(x)\,\widehat{C}^{-1}&=& \eta(x)\mid_{u\leftrightarrow d } \, ,
\end{eqnarray}
which originate from the charge conjugation property of the  axial-vector diquark states,
\begin{eqnarray}
\widehat{C}\left[\epsilon^{ijk}q^j C\gamma_\mu c^k\right]\widehat{C}^{-1}&=&\epsilon^{ijk}\bar{q}^j \gamma_\mu C \bar{c}^k \, .
\end{eqnarray}
The charged currents $\eta_{\mu\nu}(x)$ and $\eta(x)$ change   their  charge signs under charge conjugation. The neutral partners $\widetilde{\eta}_{\mu\nu}$ and $\widetilde{\eta}$,
\begin{eqnarray}
 \widetilde{\eta}_{\mu\nu}&=&\frac{\epsilon^{ijk}\epsilon^{imn}}{2}\left\{u^j C\gamma_\mu c^k \bar{u}^m\gamma_\nu C \bar{c}^n+d^j C\gamma_\mu c^k \bar{d}^m\gamma_\nu C \bar{c}^n \right.\nonumber\\
 &&\left.+tu^jC\gamma_\nu c^k\bar{u}^m\gamma_\mu C \bar{c}^n+td^jC\gamma_\nu c^k\bar{d}^m\gamma_\mu C \bar{c}^n \right\} \, , \\
  \widetilde{\eta}&=&\frac{\epsilon^{ijk}\epsilon^{imn}}{\sqrt{2}}\left\{u^jC\gamma_\mu c^k \bar{u}^m\gamma^\mu C \bar{c}^n+d^jC\gamma_\mu c^k \bar{d}^m\gamma^\mu C \bar{c}^n\right\} \, ,
\end{eqnarray}
are  eigenstates  of the  charge conjugation. The  currents $\widetilde{\eta}_{\mu\nu} $ and  $\eta_{\mu\nu}$ ($\widetilde{\eta} $ and  $\eta$) have the same charge conjugations. We  take the   currents $\eta(x)$, $\eta_{\mu\nu}^{t=-}(x)$ and $\eta_{\mu\nu}^{t=+}(x)$ to interpolate the scalar, axial-vector and tensor tetraquark states, respectively. The $Z_c(4020)$ and $Z_c(4025)$ are potential candidates of the axial-vector and tensor  tetraquark states,
the $Z(4050)$ and $Z(4250)$ are potential candidates of the scalar  tetraquark states. Other possibilities are not excluded.

At the hadronic side, we can insert  a complete set of intermediate hadronic states with
the same quantum numbers as the current operators $\eta_{\mu\nu}(x)$ and $\eta(x)$ into the
correlation functions $\Pi_{\mu\nu\alpha\beta}(p)$ and $\Pi(p)$ to obtain the hadronic representation
\cite{SVZ79,Reinders85}. After isolating the ground state
contributions of the scalar, axial-vector and tensor tetraquark states, we get the following results,
\begin{eqnarray}
\Pi_{\mu\nu\alpha\beta}^{J=2}(p)&=&\Pi_{J=2}(p)\left( \frac{\widetilde{g}_{\mu\alpha}\widetilde{g}_{\nu\beta}+\widetilde{g}_{\mu\beta}\widetilde{g}_{\nu\alpha}}{2}-\frac{\widetilde{g}_{\mu\nu}\widetilde{g}_{\alpha\beta}}{3}\right) +\Pi_s(p)\,g_{\mu\nu}g_{\alpha\beta}   \, , \nonumber\\
&=&\frac{\lambda_{ Z}^2}{M_{Z}^2-p^2}\left( \frac{\widetilde{g}_{\mu\alpha}\widetilde{g}_{\nu\beta}+\widetilde{g}_{\mu\beta}\widetilde{g}_{\nu\alpha}}{2}-\frac{\widetilde{g}_{\mu\nu}\widetilde{g}_{\alpha\beta}}{3}\right) +\cdots \, \, , \\
\Pi_{\mu\nu\alpha\beta}^{J=1}(p)&=&\Pi_{J=1}(p)\left( -\widetilde{g}_{\mu\alpha}p_{\nu}p_{\beta}-\widetilde{g}_{\nu\beta}p_{\mu}p_{\alpha}+\widetilde{g}_{\mu\beta}p_{\nu}p_{\alpha}+\widetilde{g}_{\nu\alpha}p_{\mu}p_{\beta}\right) +\Pi_s(p)\left(g_{\mu\alpha}g_{\nu\beta}-g_{\mu\beta}g_{\nu\alpha}\right) \, \, , \nonumber\\
&=&\frac{\lambda_{ Z}^2}{M_{Z}^2-p^2}\left( -\widetilde{g}_{\mu\alpha}p_{\nu}p_{\beta}-\widetilde{g}_{\nu\beta}p_{\mu}p_{\alpha}+\widetilde{g}_{\mu\beta}p_{\nu}p_{\alpha}+\widetilde{g}_{\nu\alpha}p_{\mu}p_{\beta}\right) +\cdots \, \, , \\
\Pi^{J=0}(p)&=&\Pi_{J=0}(p)=\frac{\lambda_{ Z}^2}{M_{Z}^2-p^2} +\cdots \, \, ,
\end{eqnarray}
where the notation $\widetilde{g}_{\mu\nu}=g_{\mu\nu}-\frac{p_\mu p_\nu}{p^2}$, the components  $\Pi_s(p)$ are irrelevant in the present analysis \cite{WangHcHb},
we add the superscripts and subscripts $J=2,\,1,\,0$ to denote the total angular momentum. The pole residues  $\lambda_{Z}$ are defined by
\begin{eqnarray}
 \langle 0|\eta^{t=+}_{\mu\nu}(0)|Z_{J=2}(p)\rangle &=& \lambda_{Z} \, \varepsilon_{\mu\nu} \, , \nonumber\\
 \langle 0|\eta^{t=-}_{\mu\nu}(0)|Z_{J=1}(p)\rangle &=& \lambda_{Z} \,\left(\varepsilon_{\mu}p_{\nu}-\varepsilon_{\nu}p_{\mu} \right)\, , \nonumber\\
  \langle 0|\eta(0)|Z_{J=0}(p)\rangle &=& \lambda_{Z} \, ,
\end{eqnarray}
the $\varepsilon_{\mu\nu}$ and $\varepsilon_\mu$ are the polarization vectors of the tensor and axial-vector tetraquark states respectively  with the following properties,
 \begin{eqnarray}
 \sum_{\lambda}\varepsilon^*_{\alpha\beta}(\lambda,p)\varepsilon_{\mu\nu}(\lambda,p)
 &=&\frac{\widetilde{g}_{\alpha\mu}\widetilde{g}_{\beta\nu}+\widetilde{g}_{\alpha\nu}\widetilde{g}_{\beta\mu}}{2}-\frac{\widetilde{g}_{\alpha\beta}\widetilde{g}_{\mu\nu}}{3}\, , \nonumber \\
\sum_{\lambda}\varepsilon^*_{\mu}(\lambda,p)\varepsilon_{\nu}(\lambda,p)&=&-\widetilde{g}_{\mu\nu} \, .
 \end{eqnarray}
The tensor current $\eta_{\alpha\beta}^{t=-}$ has no coupling with the $J^P=2^+$ (or tensor) tetraquark states, as the Lorentz indexes $\alpha$ and $\beta$ are antisymmetric.

 In the following,  we briefly outline  the operator product expansion for the correlation functions $\Pi_{\mu\nu\alpha\beta}(p)$ and $\Pi(p)$ in perturbative QCD.  We contract the $u$, $d$ and $c$ quark fields in the correlation functions
$\Pi_{\mu\nu\alpha\beta}(p)$ and $\Pi(p)$ with Wick theorem, and obtain the results:
\begin{eqnarray}
\Pi_{\mu\nu\alpha\beta}(p)&=&\frac{i\epsilon^{ijk}\epsilon^{imn}\epsilon^{i^{\prime}j^{\prime}k^{\prime}}\epsilon^{i^{\prime}m^{\prime}n^{\prime}}}{2}\int d^4x e^{ip \cdot x}   \nonumber\\
&&\left\{{\rm Tr}\left[ \gamma_{\mu}C^{kk^{\prime}}(x)\gamma_{\alpha} CU^{jj^{\prime}T}(x)C\right] {\rm Tr}\left[ \gamma_{\beta} C^{n^{\prime}n}(-x)\gamma_{\nu} C D^{m^{\prime}mT}(-x)C\right] \right. \nonumber\\
&&+{\rm Tr}\left[ \gamma_{\nu} C^{kk^{\prime}}(x)\gamma_{\beta} CU^{jj^{\prime}T}(x)C\right] {\rm Tr}\left[ \gamma_{\alpha} C^{n^{\prime}n}(-x)\gamma_{\mu} C D^{m^{\prime}mT}(-x)C\right] \nonumber\\
&&\pm{\rm Tr}\left[ \gamma_{\mu} C^{kk^{\prime}}(x) \gamma_{\beta} CU^{jj^{\prime}T}(x)C\right] {\rm Tr}\left[ \gamma_{\alpha} C^{n^{\prime}n}(-x) \gamma_{\nu}C D^{m^{\prime}mT}(-x)C\right] \nonumber\\
 &&\left.\pm{\rm Tr}\left[ \gamma_{\nu} C^{kk^{\prime}}(x)\gamma_{\alpha} CU^{jj^{\prime}T}(x)C\right] {\rm Tr}\left[ \gamma_{\beta} C^{n^{\prime}n}(-x)\gamma_{\mu} C D^{m^{\prime}mT}(-x)C\right] \right\} \, , \nonumber\\
 \Pi(p)&=&i\epsilon^{ijk}\epsilon^{imn}\epsilon^{i^{\prime}j^{\prime}k^{\prime}}\epsilon^{i^{\prime}m^{\prime}n^{\prime}}\int d^4x e^{ip \cdot x}   \nonumber\\
&&{\rm Tr}\left[ \gamma_{\mu}C^{kk^{\prime}}(x)\gamma_{\alpha} CU^{jj^{\prime}T}(x)C\right] {\rm Tr}\left[ \gamma^{\alpha} C^{n^{\prime}n}(-x)\gamma^{\mu} C D^{m^{\prime}mT}(-x)C\right]   \, ,
\end{eqnarray}
where the $\pm$ correspond to $t=\pm$ respectively,
 the $U_{ij}(x)$, $D_{ij}(x)$ and $C_{ij}(x)$ are the full $u$, $d$ and $c$ quark propagators respectively (the $U_{ij}(x)$ and $D_{ij}(x)$ can be written as $S_{ij}(x)$ for simplicity),
 \begin{eqnarray}
S_{ij}(x)&=& \frac{i\delta_{ij}\!\not\!{x}}{ 2\pi^2x^4}-\frac{\delta_{ij}\langle
\bar{q}q\rangle}{12} -\frac{\delta_{ij}x^2\langle \bar{q}g_s\sigma Gq\rangle}{192} -\frac{ig_sG^{a}_{\alpha\beta}t^a_{ij}(\!\not\!{x}
\sigma^{\alpha\beta}+\sigma^{\alpha\beta} \!\not\!{x})}{32\pi^2x^2} -\frac{i\delta_{ij}x^2\!\not\!{x}g_s^2\langle \bar{q} q\rangle^2}{7776}\nonumber\\
&&  -\frac{\delta_{ij}x^4\langle \bar{q}q \rangle\langle g_s^2 GG\rangle}{27648} -\frac{1}{8}\langle\bar{q}_j\sigma^{\mu\nu}q_i \rangle \sigma_{\mu\nu}-\frac{1}{4}\langle\bar{q}_j\gamma^{\mu}q_i\rangle \gamma_{\mu }+\cdots \, ,
\end{eqnarray}
\begin{eqnarray}
C_{ij}(x)&=&\frac{i}{(2\pi)^4}\int d^4k e^{-ik \cdot x} \left\{
\frac{\delta_{ij}}{\!\not\!{k}-m_c}
-\frac{g_sG^n_{\alpha\beta}t^n_{ij}}{4}\frac{\sigma^{\alpha\beta}(\!\not\!{k}+m_c)+(\!\not\!{k}+m_c)
\sigma^{\alpha\beta}}{(k^2-m_c^2)^2}\right.\nonumber\\
&&\left. +\frac{g_s D_\alpha G^n_{\beta\lambda}t^n_{ij}(f^{\lambda\beta\alpha}+f^{\lambda\alpha\beta}) }{3(k^2-m_c^2)^4}-\frac{g_s^2 (t^at^b)_{ij} G^a_{\alpha\beta}G^b_{\mu\nu}(f^{\alpha\beta\mu\nu}+f^{\alpha\mu\beta\nu}+f^{\alpha\mu\nu\beta}) }{4(k^2-m_c^2)^5}+\cdots\right\} \, ,\nonumber\\
f^{\lambda\alpha\beta}&=&(\!\not\!{k}+m_c)\gamma^\lambda(\!\not\!{k}+m_c)\gamma^\alpha(\!\not\!{k}+m_c)\gamma^\beta(\!\not\!{k}+m_c)\, ,\nonumber\\
f^{\alpha\beta\mu\nu}&=&(\!\not\!{k}+m_c)\gamma^\alpha(\!\not\!{k}+m_c)\gamma^\beta(\!\not\!{k}+m_c)\gamma^\mu(\!\not\!{k}+m_c)\gamma^\nu(\!\not\!{k}+m_c)\, ,
\end{eqnarray}
and  $t^n=\frac{\lambda^n}{2}$, the $\lambda^n$ is the Gell-Mann matrix,  $D_\alpha=\partial_\alpha-ig_sG^n_\alpha t^n$ \cite{Reinders85}, then compute  the integrals both in the coordinate and momentum spaces to obtain the correlation functions $\Pi_{\mu\nu\alpha\beta}(p)$ and $\Pi(p)$ therefore the QCD spectral densities.
In Eq.(16), we retain the terms $\langle\bar{q}_j\sigma_{\mu\nu}q_i \rangle$ and $\langle\bar{q}_j\gamma_{\mu}q_i\rangle$ originate from the Fierz re-arrangement of the $\langle q_i \bar{q}_j\rangle$ to  absorb the gluons  emitted from the heavy quark lines to form $\langle\bar{q}_j g_s G^a_{\alpha\beta} t^a_{mn}\sigma_{\mu\nu} q_i \rangle$ and $\langle\bar{q}_j\gamma_{\mu}q_ig_s D_\nu G^a_{\alpha\beta}t^a_{mn}\rangle$ so as to extract the mixed condensate and four-quark condensates $\langle\bar{q}g_s\sigma G q\rangle$ and $g_s^2\langle\bar{q}q\rangle^2$, respectively. One can consult Ref.\cite{WangHuangTao} for some technical details in the operator product expansion.

 Once the analytical expressions  are obtained,  we can take the
quark-hadron duality below the continuum thresholds  $s_0$ and perform Borel transform  with respect to
the variable $P^2=-p^2$ to obtain  the following QCD sum rules:
\begin{eqnarray}
\lambda^2_{Z}\, \exp\left(-\frac{M^2_{Z}}{T^2}\right)= \int_{4m_c^2}^{s_0} ds\, \rho(s) \, \exp\left(-\frac{s}{T^2}\right) \, ,
\end{eqnarray}
where
\begin{eqnarray}
\rho(s)&=&\rho_{0}(s)+\rho_{3}(s) +\rho_{4}(s)+\rho_{5}(s)+\rho_{6}(s)+\rho_{7}(s) +\rho_{8}(s)+\rho_{10}(s)\, ,
\end{eqnarray}
  the subscripts $0$, $3$, $4$, $5$, $6$, $7$, $8$ and $10$ denote the dimensions of the vacuum condensates in the operator product expansion, the $T^2$ denotes the Borel parameter. The explicit expressions of the QCD spectral densities $\rho_i(s)$ are given in the appendix.
 In this article, we carry out the
operator product expansion to the vacuum condensates  up to dimension-10 and discard the perturbative corrections. Furthermore, we  assume  vacuum saturation for the  higher dimension vacuum condensates.  The terms associate with $\frac{1}{T^2}$, $\frac{1}{T^4}$, $\frac{1}{T^6}$ in the QCD spectral densities $\rho_{i}$, $i=7,8,10$, manifest themselves at small values of the $T^2$, we have to choose large values of the $T^2$ to warrant convergence of the operator product expansion and appearance of the Borel platforms. In the Borel windows, the higher dimension vacuum condensates  play a less important role.
In summary, the higher dimension vacuum condensates play an important role in determining the Borel windows therefore the ground state  masses and pole residues, though they maybe play a less important role in the Borel windows. We should take them into account consistently.

 Differentiate   Eq.(18) with respect to  $\frac{1}{T^2}$, then eliminate the
 pole residues $\lambda_{Z}$, we obtain the QCD sum rules for
 the masses of the scalar, axial-vector and tensor    tetraquark states,
 \begin{eqnarray}
 M^2_{Z}= \frac{\int_{4m_c^2}^{s_0} ds\frac{d}{d \left(-1/T^2\right)}\rho(s)\exp\left(-\frac{s}{T^2}\right)}{\int_{4m_c^2}^{s_0} ds \rho(s)\exp\left(-\frac{s}{T^2}\right)}\, .
\end{eqnarray}

\section{Numerical results and discussions}
The vacuum condensates are taken to be the standard values
$\langle\bar{q}q \rangle=-(0.24\pm 0.01\, \rm{GeV})^3$,
$\langle\bar{q}g_s\sigma G q \rangle=m_0^2\langle \bar{q}q \rangle$,
$m_0^2=(0.8 \pm 0.1)\,\rm{GeV}^2$, $\langle \frac{\alpha_s
GG}{\pi}\rangle=(0.33\,\rm{GeV})^4 $    at the energy scale  $\mu=1\, \rm{GeV}$
\cite{SVZ79,Reinders85,Ioffe2005}.
The quark condensate and mixed quark condensate evolve with the   renormalization group equation,
$\langle\bar{q}q \rangle(\mu)=\langle\bar{q}q \rangle(Q)\left[\frac{\alpha_{s}(Q)}{\alpha_{s}(\mu)}\right]^{\frac{4}{9}}$ and
 $\langle\bar{q}g_s \sigma Gq \rangle(\mu)=\langle\bar{q}g_s \sigma Gq \rangle(Q)\left[\frac{\alpha_{s}(Q)}{\alpha_{s}(\mu)}\right]^{\frac{2}{27}}$.

In the article, we take the $\overline{MS}$ mass $m_{c}(m_c)=(1.275\pm0.025)\,\rm{GeV}$
 from the Particle Data Group \cite{PDG}, and take into account
the energy-scale dependence of  the $\overline{MS}$ mass from the renormalization group equation,
\begin{eqnarray}
m_c(\mu)&=&m_c(m_c)\left[\frac{\alpha_{s}(\mu)}{\alpha_{s}(m_c)}\right]^{\frac{12}{25}} \, ,\nonumber\\
\alpha_s(\mu)&=&\frac{1}{b_0t}\left[1-\frac{b_1}{b_0^2}\frac{\log t}{t} +\frac{b_1^2(\log^2{t}-\log{t}-1)+b_0b_2}{b_0^4t^2}\right]\, ,
\end{eqnarray}
  where $t=\log \frac{\mu^2}{\Lambda^2}$, $b_0=\frac{33-2n_f}{12\pi}$, $b_1=\frac{153-19n_f}{24\pi^2}$, $b_2=\frac{2857-\frac{5033}{9}n_f+\frac{325}{27}n_f^2}{128\pi^3}$,  $\Lambda=213\,\rm{MeV}$, $296\,\rm{MeV}$  and  $339\,\rm{MeV}$ for the flavors  $n_f=5$, $4$ and $3$, respectively  \cite{PDG}.

In the conventional QCD sum rules \cite{SVZ79,Reinders85}, there are
two criteria (pole dominance and convergence of the operator product
expansion) for choosing  the Borel parameter $T^2$ and threshold
parameter $s_0$. The light tetraquark states cannot satisfy the two
criteria,  though it is not an indication of non-existence of the
light tetraquark states (For detailed discussions about this
subject, one can consult Ref.\cite{Wang-NPA}). We impose
the two criteria on the hidden charmed tetraquark states to choose the Borel
parameter $T^2$ and threshold parameter $s_0$. Furthermore, we
  take it for granted that the energy gap between the ground
states and the first radial excited states is about $0.5\,\rm{GeV}$.

In Fig.1,  the masses of the scalar ($J=0$), axial-vector ($J=1$) and tensor ($J=2$) tetraquark states  are plotted   with variations of the  Borel parameters $T^2$, energy scales $\mu$, and continuum  threshold parameters $s_0$, where $s_0^{J=2}=s_0^{J=1}=20.5\,\rm{GeV}^2$, $s_0^{J=0}=18.5\,\rm{GeV}^2$, $s_0^{Z(4050)}=21.0\,\rm{GeV}^2$, and $s_0^{Z(4250)}=22.5\,\rm{GeV}^2$. From the figure, we can see that the masses decrease monotonously with increase of the energy scales. The parameters $s_0^{J=2}=s_0^{J=1}=20.5\,\rm{GeV}^2$ and $\mu = 1.8\,\rm{GeV}$ can reproduce the experimental values of the masses of the $Z_c(4020)$ and $Z_c(4025)$ in the cases of $J^{PC}=1^{+-}$ or $2^{++}$ assignments.
In Ref.\cite{Wang1311},  we observe that the predictions based on the QCD sum rules  disfavor assigning  the $Z_c(4020)$ and $Z_c(4025)$ as the
diquark-antidiquark type vector tetraquark states.

\begin{figure}
\centering
\includegraphics[totalheight=6cm,width=7cm]{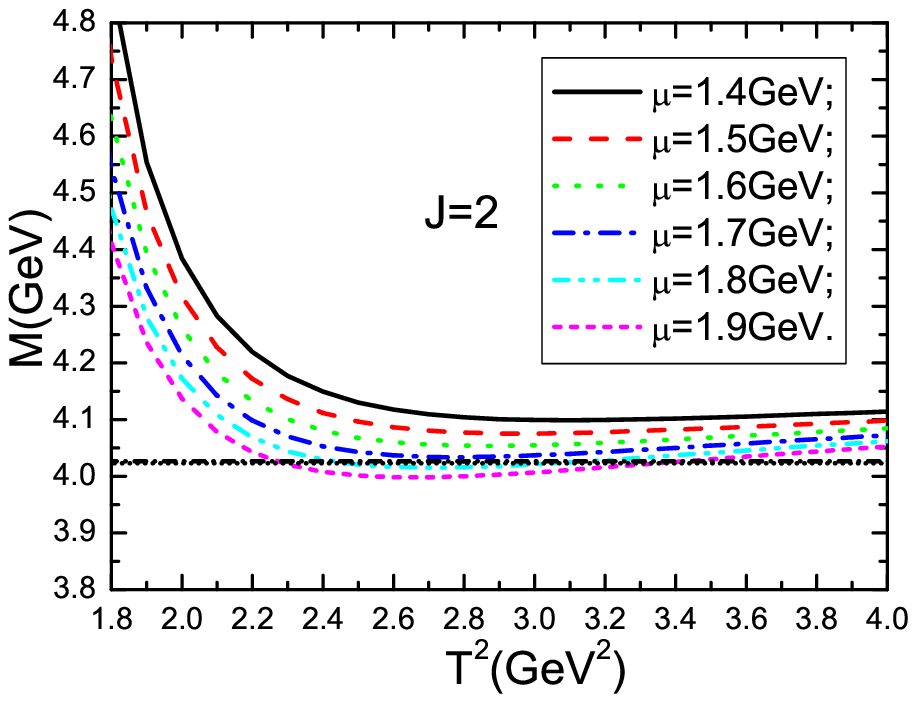}
\includegraphics[totalheight=6cm,width=7cm]{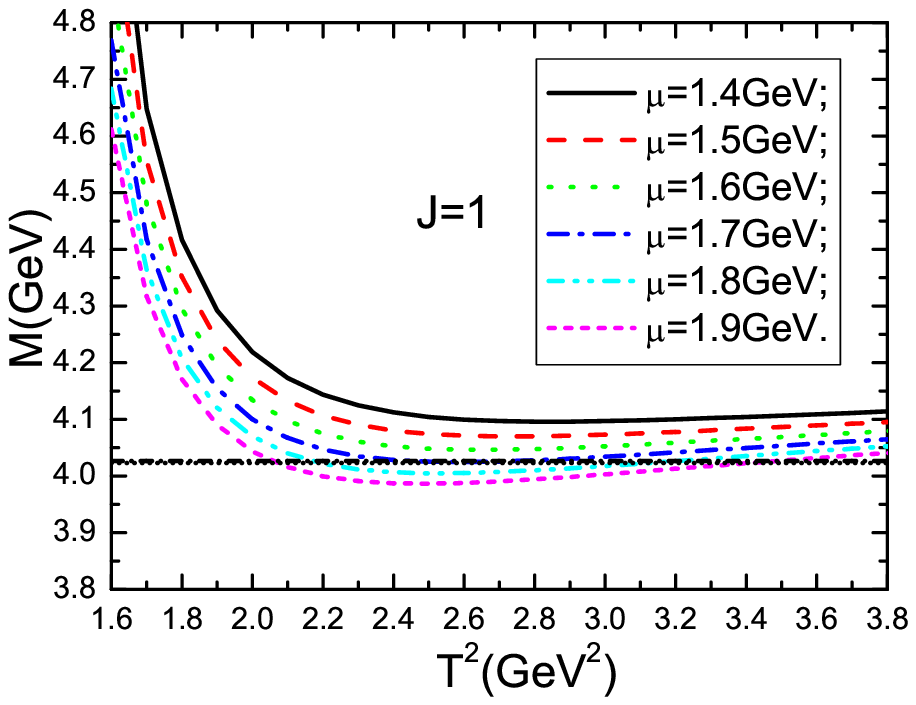}
\includegraphics[totalheight=6cm,width=7cm]{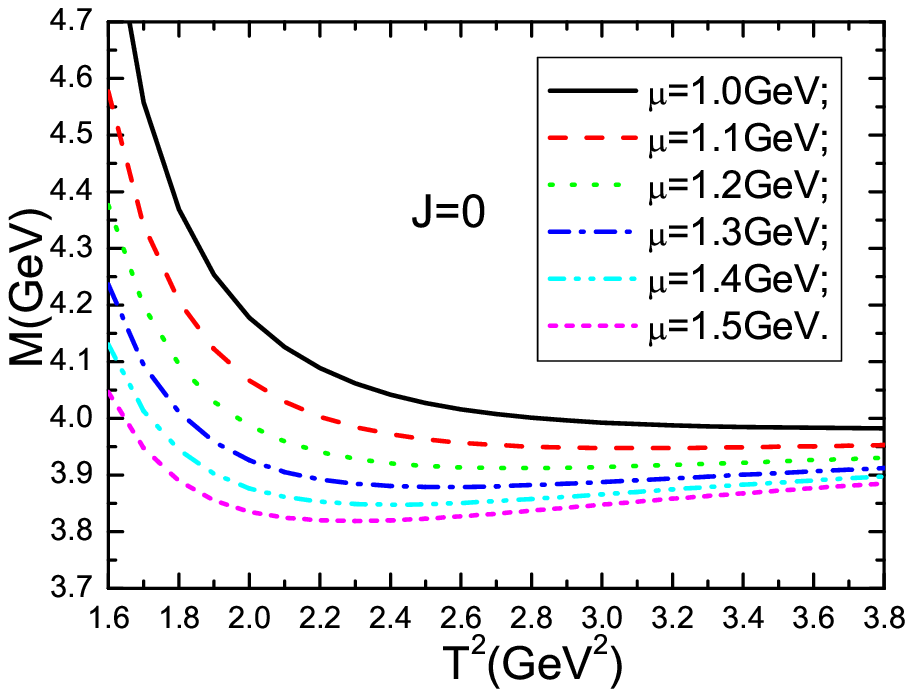}
\includegraphics[totalheight=6cm,width=7cm]{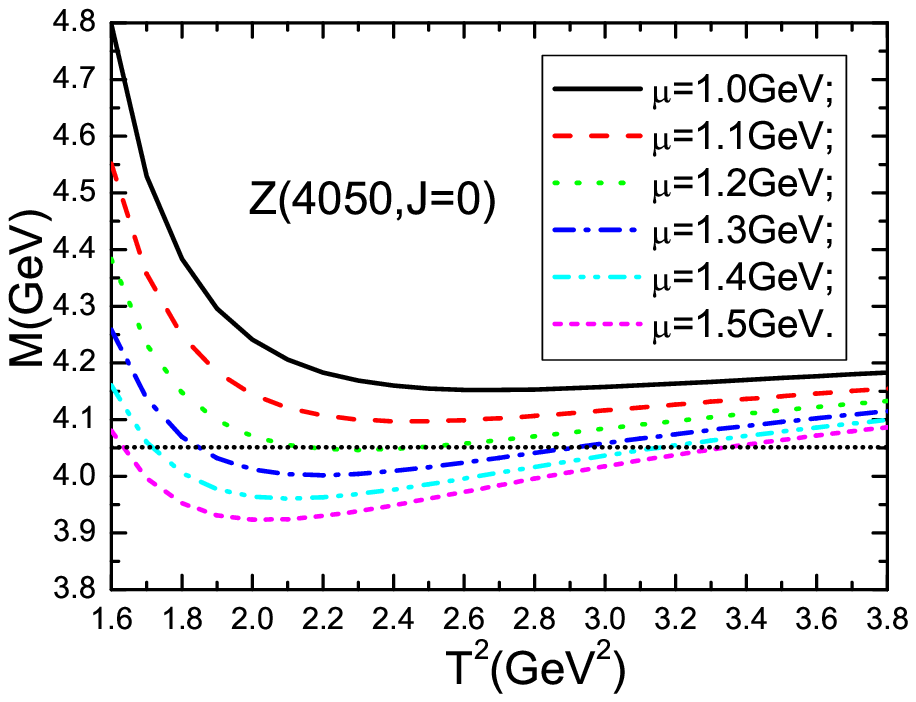}
\includegraphics[totalheight=6cm,width=7cm]{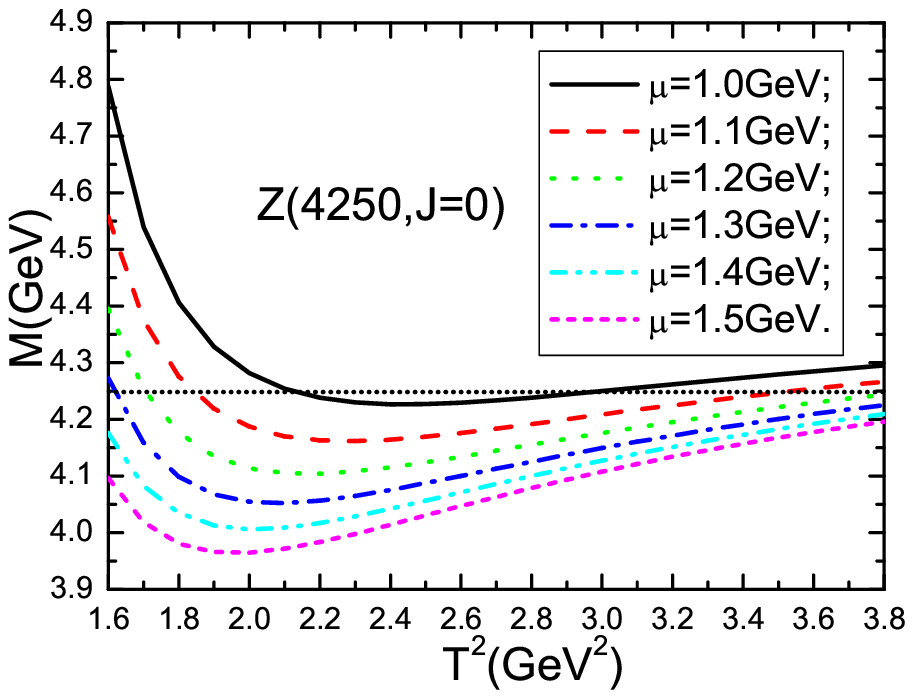}
  \caption{ The masses  with variations of the  Borel parameters $T^2$ and energy scales $\mu$ for the $J=2,1,0$ tetraquark states. The horizontal lines denote the experimental values of the masses of the $Z_c(4020)$ and $Z_c(4025)$ in the case of $J=1$ and $J=2$;  the experimental values of the masses of the $Z(4050)$ and $Z(4250)$ in the case of $Z(4050,J=0)$ and $Z(4250,J=0)$. In the figures denoted by the $Z(4050,J=0)$ and $Z(4250,J=0)$, we take the $Z(4050)$ and $Z(4250)$ as the $J=0$ tetraquark states,
  choose the continuum threshold parameters $s_0^{Z(4050)}=(M_{Z(4050)}+0.5)^2\approx21.0\,\rm{GeV}^2$ and $s_0^{Z(4250)}=(M_{Z(4250)}+0.5)^2\approx22.5\,\rm{GeV}^2$. }
\end{figure}

In Ref.\cite{WangHuangTao}, we observe that  the  energy scale $\mu=1.5\,\rm{GeV}$ is the lowest energy scale to reproduce the experimental values of the masses of the $X(3872)$ and $Z_c(3900)$ or $Z_c(3885)$. In Ref.\cite{Wang1311}, we  suggest  a  formula to estimate the energy scales of the QCD spectral densities in the QCD sum rules for the  hidden charmed tetraquark states,
 $ \mu=\sqrt{M^2_{X/Y/Z}-(2{\mathbb{M}}_c)^2}$, with the effective $c$-quark mass ${\mathbb{M}}_c=1.8\,\rm{GeV}$.
 The heavy tetraquark system could be described
by a double-well potential with two light quarks $q^{\prime}\bar{q}$ lying in the two wells respectively.
   In the heavy quark limit, the $c$ (and $b$) quark can be taken as a static well potential,
which binds the light quark $q^{\prime}$ to form a diquark in the color antitriplet channel or binds the light antiquark $\bar{q}$ to form a meson in the color singlet channel (or a meson-like state in the color octet  channel). Then the heavy tetraquark states  are characterized by the effective heavy quark masses ${\mathbb{M}}_Q$ (or constituent quark masses) and the virtuality $\sqrt{M^2_{X/Y/Z}-(2{\mathbb{M}}_Q)^2}$ (or bound energy not as robust). It is natural to take the energy  scale $\mu=$virtuality.
 The resulting  energy scales are  $\mu=1.5\,\rm{GeV}$ for the $Z_c(3900)$ and $X(3872)$ \cite{WangHuangTao}, $\mu=3.0\,\rm{GeV}$ for the $Y(4660)$ \cite{Wang1311}, $\mu=1.8\,\rm{GeV}$ for the $Z_c(4020)$ and $Z_c(4025)$ in the present case.

If  the $Z(4050)$ and $Z(4250)$   are scalar tetraquark states,  the energy scale $\mu=1.8\,\rm{GeV}$ is too large, see Fig.1.  \\
$\bullet$ If the $Z(4050)$ is a scalar tetraquark state, the energy scale $\mu=1.2\,\rm{GeV}$ is the optimal energy scale.\\
$\bullet$ If the $Z(4250)$ is a scalar tetraquark state, the energy scale $\mu=1.0\,\rm{GeV}$ is the optimal energy scale.\\
In the two cases, the energy scales $\mu\ll 1.8\,\rm{GeV}$, it is very odd that the energy scales of the QCD spectral densities of the  $Z(4050)$ and $Z(4250)$ are much smaller than that of the  $Z_c(4020)$ and $Z_c(4025)$.
We can draw the conclusion tentatively that the predictions based on the QCD sum rules disfavor assigning the $Z(4050)$ and $Z(4250)$ as the   $C\gamma_\mu-C\gamma^\mu$ type scalar tetraquark states.

In Refs.\cite{Wang4250,Wang4250-EPJC}, we obtain the masses    $(4.36 \pm 0.18)\, \rm{GeV}$,  $(4.37 \pm 0.18)\, \rm{GeV}$ and $(4.56 \pm 0.14)\,\rm{GeV}$ for the  $C\gamma_\mu-C\gamma^\mu$,  $C\gamma_5-C\gamma_5$ and $C\gamma_\mu\gamma_5-C\gamma^\mu\gamma_5$ type   scalar $c\bar{c}c\bar{d}$ tetraquark states respectively using  the QCD sum rules. In calculations, we take the mass $m_c({\mu=1\,\rm{GeV}}) = (1.35 \pm 0.10)\,\rm{GeV}$, the energy scale $\mu=1\,\rm{GeV}$ is much smaller than that determined by  the formula $ \mu=\sqrt{M^2_{X/Y/Z}-(2{\mathbb{M}}_c)^2}$. The masses of the $C\gamma_\mu\gamma_5-C\gamma^\mu\gamma_5$ type   scalar  tetraquark states are much larger than that of the $C\gamma_\mu-C\gamma^\mu$ and  $C\gamma_5-C\gamma_5$ type scalar  tetraquark states \cite{Wang4250,Wang4250-EPJC}. If the $Z(4050)$ and $Z(4250)$ are scalar tetraquark states, they may be the $C\gamma_\mu\gamma_5-C\gamma^\mu\gamma_5$ or $C-C$ type.

 Under the Fierz re-arrangement, we can rearrange the diquark-antidiquark type  currents into the color singlet-singlet type + color octet-octet type currents \cite{Ni-Da}, the energy scale  formula can be applied  to  study  the molecular states  in the QCD sum rules \cite{Wang4140}.

We take the energy scales  as $\mu=1.8\,\rm{GeV}$, $1.8\,\rm{GeV}$ and $1.4\,\rm{GeV}$ for the $J=2$, $1$  and  $0$ tetraquark states, respectively.
In Fig.2,  the contributions of the pole terms are plotted with
variations of the threshold parameters $s_0$ and Borel parameters $T^2$.
In calculations, we observe
 that the values  $s_0^{J=2}\leq 18.5 \, \rm{GeV}^2$, $s_0^{J=1}\leq 18.5 \, \rm{GeV}^2$ and $ s_0^{J=0}\leq16.5 \, \rm{GeV}^2$   are too small to satisfy the pole dominance condition and result in reasonable Borel windows.
In Fig.3,  the contributions of different terms in the
operator product expansion are plotted with variations of the Borel parameters  $T^2$ for the threshold parameters $s_0^{J=2}=s_0^{J=1}=20.5\,\rm{GeV}^2$,  $s_0^{J=0}=18.5\,\rm{GeV}^2$.  The contributions of the vacuum condensates of dimensions 3, 5, 6, 8, 10 change quickly with variations of the Borel parameters
at the regions $T^2< 2.4\,\rm{GeV}^2$ and  $2.0\,\rm{GeV}^2$ for the $J=2$   and  $0$ tetraquark states,  respectively,  the contributions of the vacuum condensates of dimensions 3, 5 change quickly with variations of the Borel parameter
at the region $T^2< 2.4\,\rm{GeV}^2$ for the $J=1$    tetraquark state, which does  not warrant
platforms for the masses.

In this article,   the Borel parameters are chosen as
 $T^2=(2.6-2.8)\,\rm{GeV}^2$, $(2.8-3.2)\,\rm{GeV}^2$ and $(2.0-2.4)\,\rm{GeV}^2$    for the $J=2$, $1$  and  $0$ tetraquark states,  respectively,
 the continuum  threshold parameters are chosen as
 $s_0=(20.5\pm1.0)\,\rm{GeV}^2$, $(20.5\pm1.0)\,\rm{GeV}^2$ and $(18.5\pm1.0)\,\rm{GeV}^2$    for the $J=2$, $1$  and  $0$ tetraquark states, respectively.
Then the convergent
behavior in the operator product  expansion is very good. Such Borel parameters and threshold parameters  can also lead to  analogous pole contributions.
The Borel parameters, continuum threshold parameters and the pole contributions are shown explicitly in Table 1. The two criteria (pole dominance and convergence of the operator product expansion) of the QCD sum rules are fully satisfied, so we expect to make reasonable predictions.

\begin{table}
\begin{center}
\begin{tabular}{|c|c|c|c|c|c|c|c|}\hline\hline
   $J^{PC}$   & $T^2 (\rm{GeV}^2)$ & $s_0 (\rm{GeV}^2)$   & pole         & $M_{Z}(\rm{GeV})$         & $\lambda_{Z}$ \\ \hline
   $2^{++}$   & $2.6-3.0$          & $20.5\pm1.0$         & $(50-73)\%$  & $4.02^{+0.09}_{-0.09}$    & $3.97^{+0.58}_{-0.52}\times10^{-2}\rm{GeV}^5$       \\ \hline
   $1^{+-}$   & $2.8-3.2$          & $20.5\pm1.0$         & $(51-72)\%$  & $4.02^{+0.07}_{-0.08}$    & $0.80^{+0.11}_{-0.11}\times10^{-2}\rm{GeV}^4$    \\ \hline
   $0^{++}$   & $2.0-2.4$          & $18.5\pm1.0$         & $(50-78)\%$  & $3.85^{+0.15}_{-0.09}$    & $3.15^{+0.75}_{-0.53}\times10^{-2}\rm{GeV}^5$    \\ \hline
 \hline
\end{tabular}
\end{center}
\caption{ The Borel parameters, continuum threshold parameters, pole contributions, masses and pole residues of the scalar, axial-vector and tensor  tetraquark states. }
\end{table}

\begin{figure}
\centering
\includegraphics[totalheight=6cm,width=7cm]{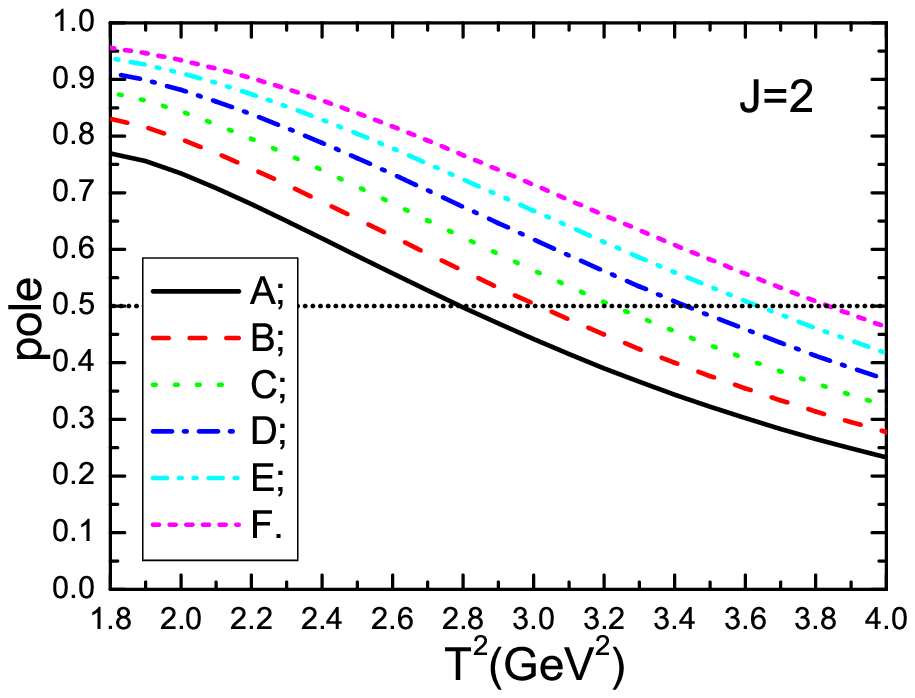}
\includegraphics[totalheight=6cm,width=7cm]{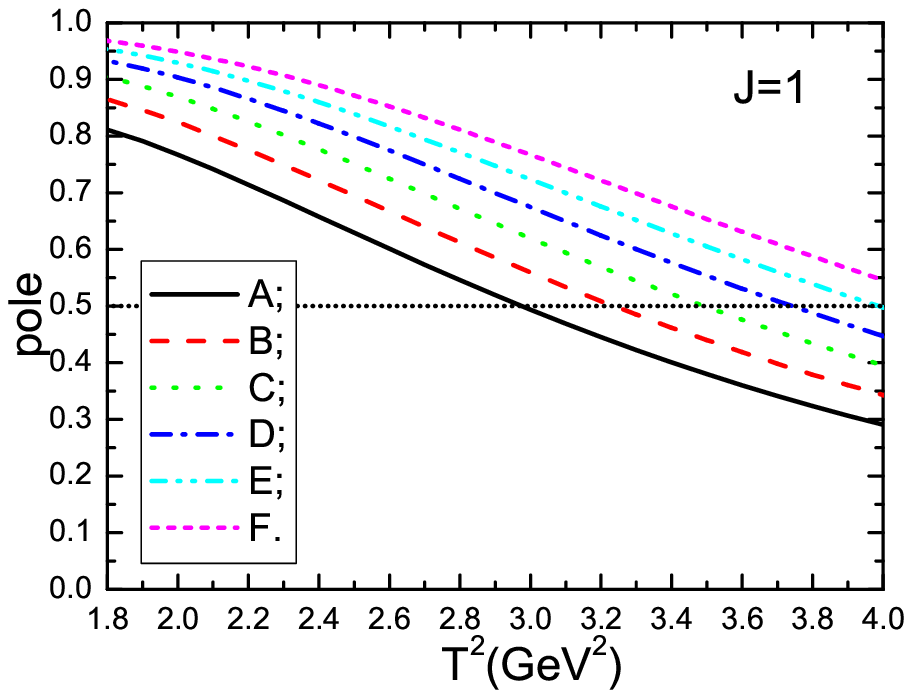}
\includegraphics[totalheight=6cm,width=7cm]{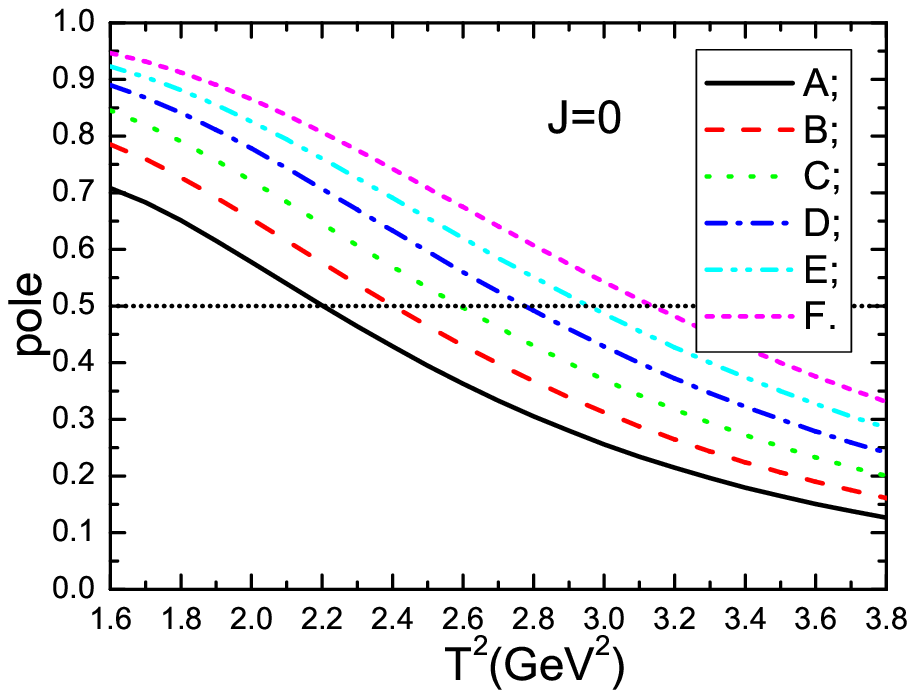}
  \caption{ The pole contributions   with variations of the  Borel parameters $T^2$ and threshold parameters $s_0$, where the $A$, $B$, $C$, $D$, $E$ and $F$ denote the threshold parameters $s_0=18.5$, 19.5, 20.5, 21.5, 22.5 and $23.5\,\rm{GeV}^2$ respectively for the  $J=2,\,1$  tetraquark states; $16.5$, 17.5, 18.5, 19.5, 20.5 and $21.5\,\rm{GeV}^2$ respectively for the  $J=0$  tetraquark state. }
\end{figure}

\begin{figure}
\centering
\includegraphics[totalheight=6cm,width=7cm]{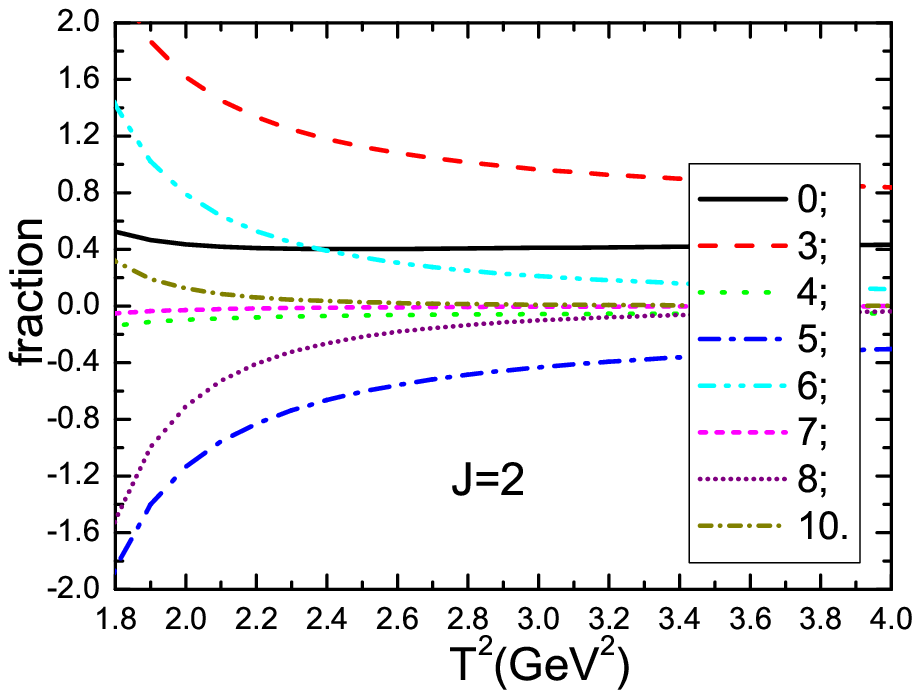}
\includegraphics[totalheight=6cm,width=7cm]{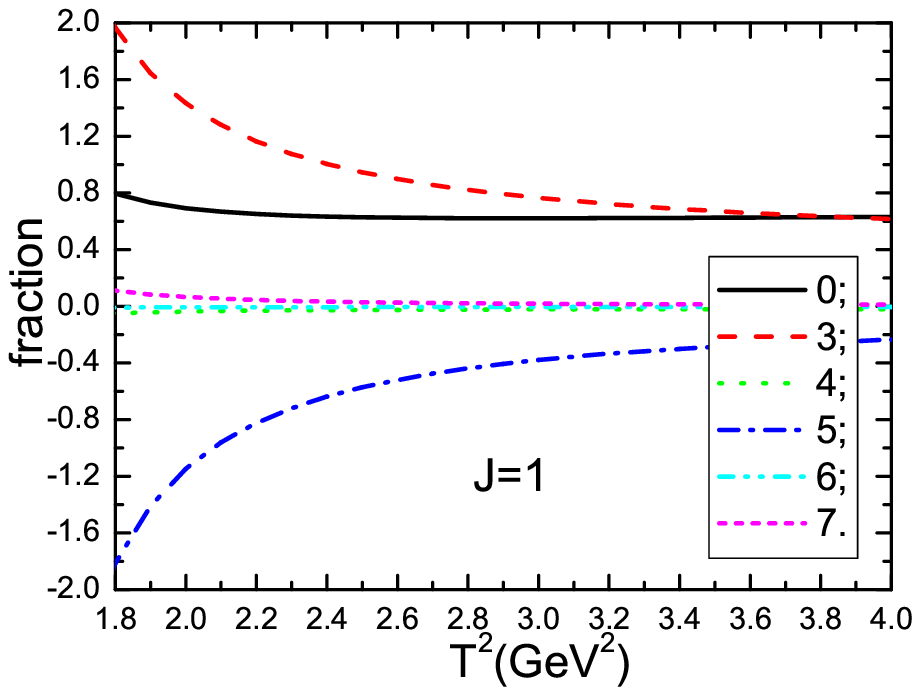}
\includegraphics[totalheight=6cm,width=7cm]{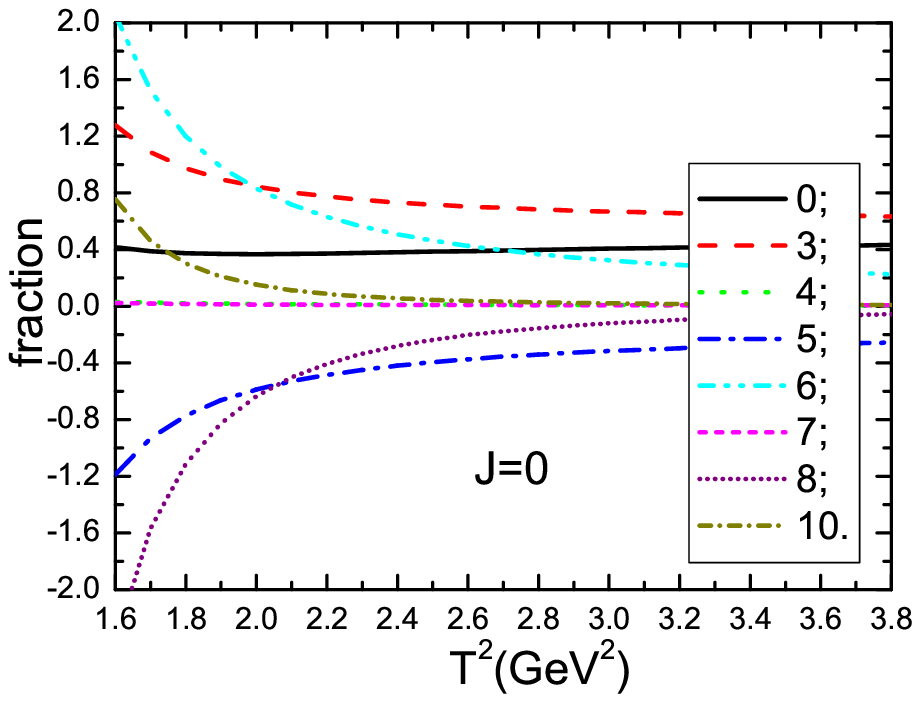}
  \caption{ The contributions of different terms in the operator product expansion  with variations of the  Borel parameters $T^2$, where the 0, 3, 4, 5, 6 ,7, 8 and 10 denote the dimensions of the vacuum condensates. }
\end{figure}

\begin{figure}
\centering
\includegraphics[totalheight=6cm,width=7cm]{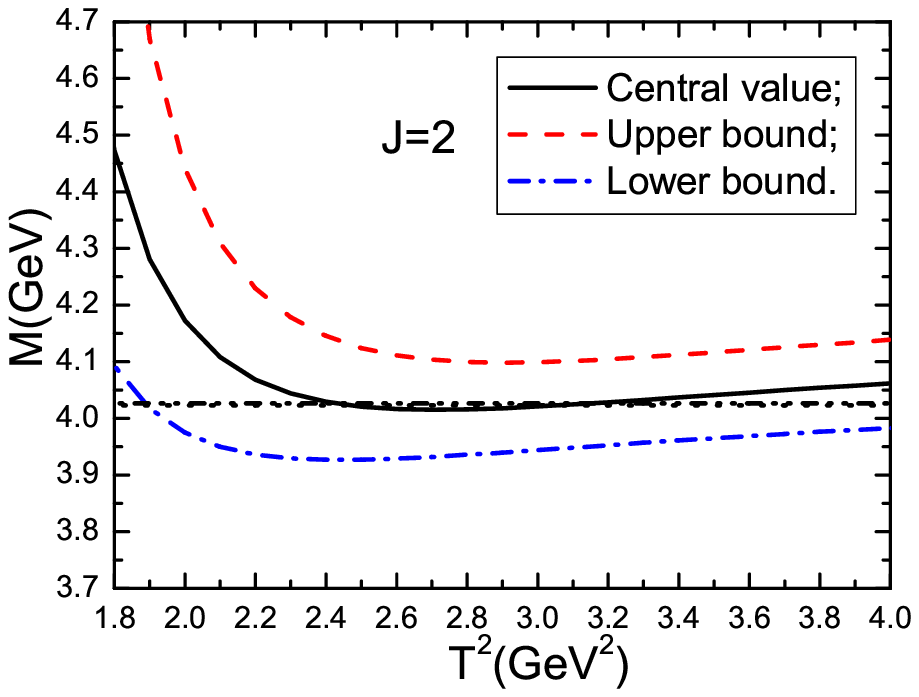}
\includegraphics[totalheight=6cm,width=7cm]{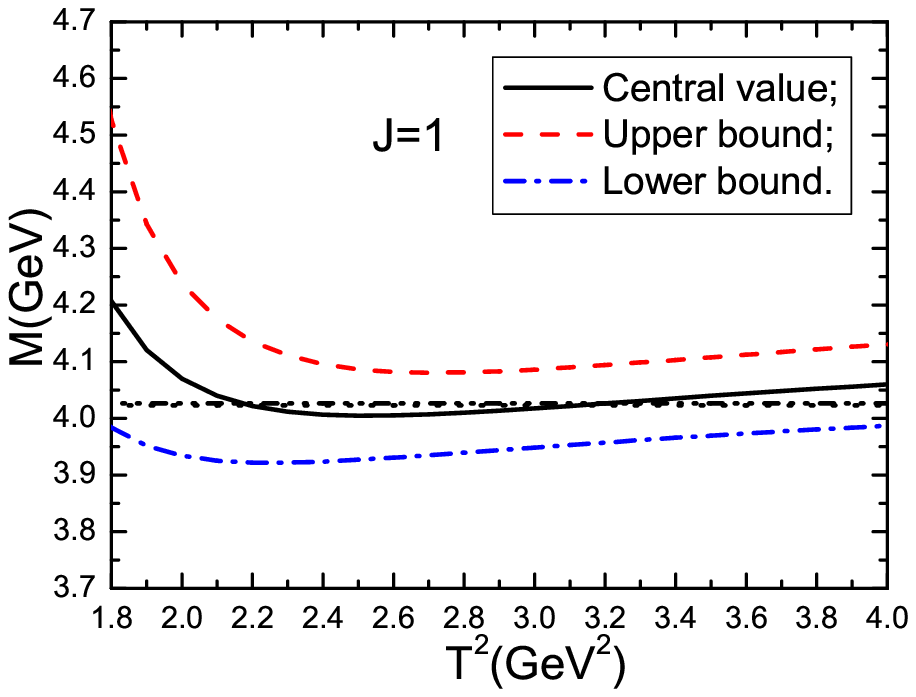}
\includegraphics[totalheight=6cm,width=7cm]{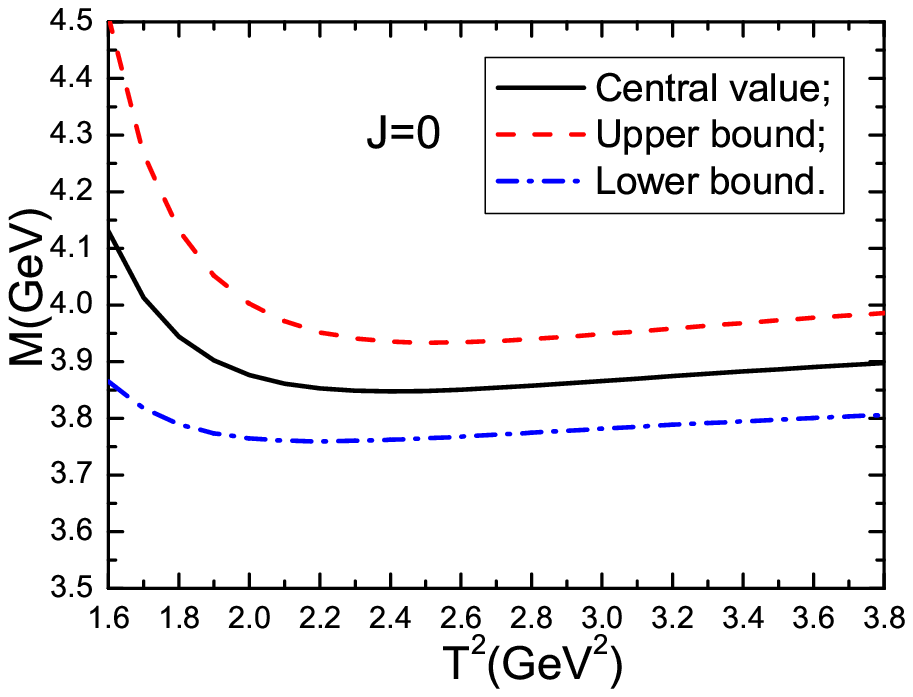}
  \caption{ The masses  with variations of the  Borel parameters $T^2$, where the horizontal lines denote the experimental values of the masses of the $Z_c(4020)$ and $Z_c(4025)$ . }
\end{figure}

\begin{figure}
\centering
\includegraphics[totalheight=6cm,width=7cm]{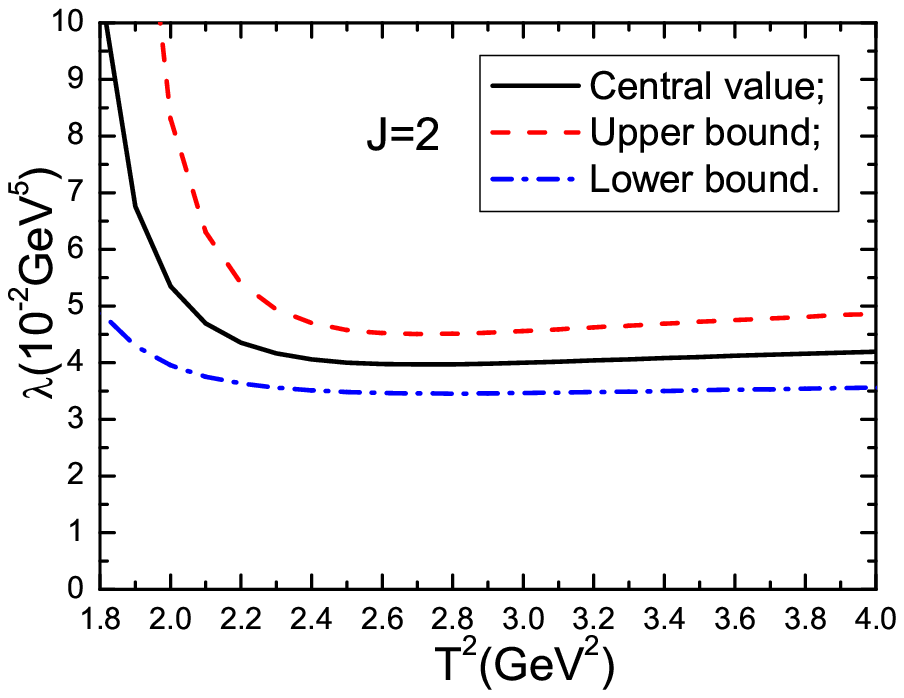}
\includegraphics[totalheight=6cm,width=7cm]{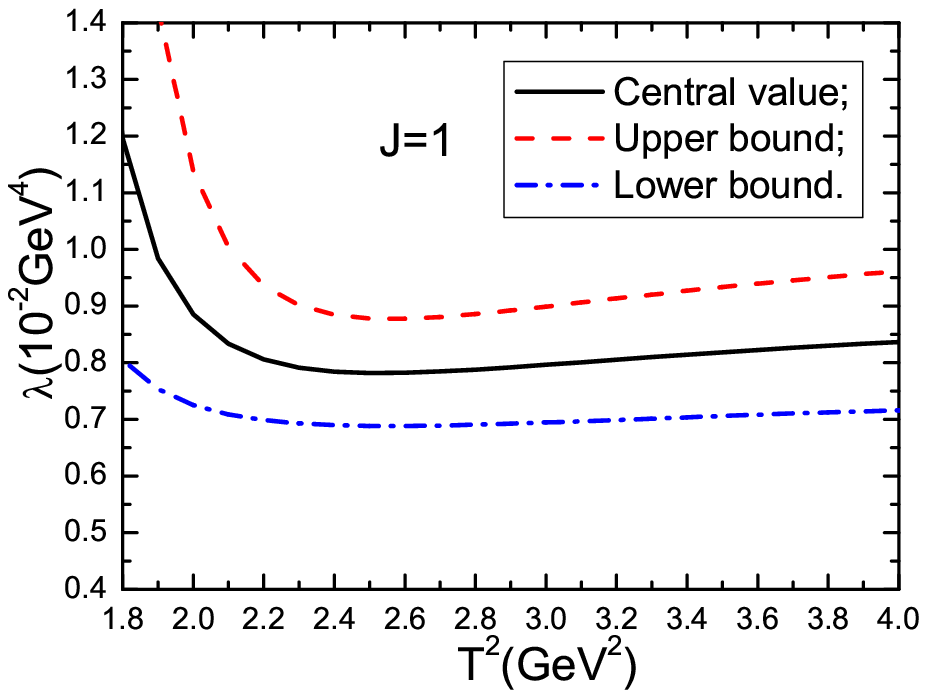}
\includegraphics[totalheight=6cm,width=7cm]{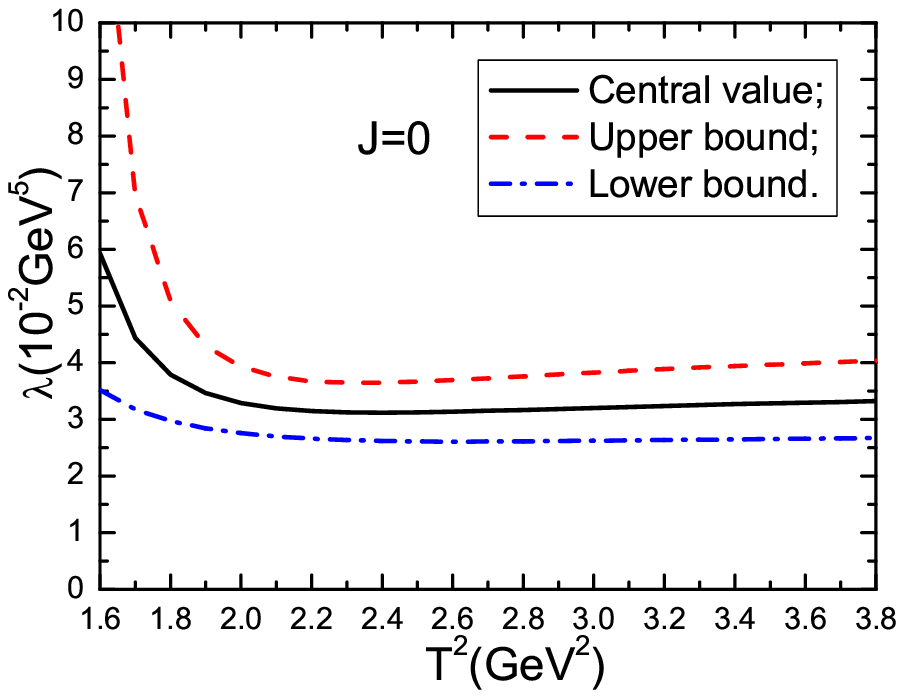}
  \caption{ The pole residues  with variations of the  Borel parameters $T^2$. }
\end{figure}

Taking into account all uncertainties of the input parameters,
finally we obtain the values of the masses and pole residues of
 the   scalar, axial-vector and tensor tetraquark states, which are  shown explicitly in Figs.4-5 and Table 1.
The predictions  $M_{J=2} =\left(4.02^{+0.09}_{-0.09}\right)\,\rm{GeV}$, $M_{J=1} =\left(4.02^{+0.07}_{-0.08}\right)\,\rm{GeV}$ are consistent with the experimental values $M_{Z_c(4025)}=(4026.3\pm2.6\pm3.7)\,\rm{MeV}$, $M_{Z_c(4020)}=(4022.9\pm 0.8\pm 2.7)\,\rm{MeV}$ from the BESIII collaboration \cite{BES1308,BES1309}.   The present predictions favor assigning the $Z_c(4020)$ and $Z_c(4025)$   as the $J^{PC}=1^{+-}$ and $2^{++}$   diquark-antidiquark type tetraquark states.  More experimental data on the spin and parity are stilled needed to identify the $Z_c(4020)$ and $Z_c(4025)$, while the $Z(4050)$ and $Z(4250)$ still need confirmation.

In Table 1, the pole residues $\lambda_{Z(1^{+-})}=\left(0.80^{+0.11}_{-0.11}\right)\times10^{-2}\rm{GeV}^4$ and $\lambda_{Z(2^{++})}=\left(3.97^{+0.58}_{-0.52}\right)\times10^{-2}\rm{GeV}^5$ are quite different due to the definitions, see Eq.(13), although the interpolating currents $\eta_{\mu\nu}^{t=\pm}$ are similar.  The  $\lambda_{Z(1^{+-})}$ has smaller dimension of mass compared to the $\lambda_{Z(2^{++})}$, the $\lambda_{Z(0^{++})}$, $\lambda_{Z(1^{+-})}M_{Z(1^{+-})}$, $\lambda_{Z(2^{++})}$ are of the same magnitude.  The correlation function $\Pi^{J=2}_{\mu\nu\alpha\beta}(p)$ ($\Pi^{J=1}_{\mu\nu\alpha\beta}(p)$) is symmetric (anti-symmetric) in  exchanging the Lorentz indexes $\mu\leftrightarrow\nu$ or $\alpha\leftrightarrow\beta$, the vacuum condensates of the dimensions 8 and 10 and some other vacuum condensates disappear at  the QCD side of the correlation function $\Pi^{J=1}_{\mu\nu\alpha\beta}(p)$ due to the anti-symmetry property of the Lorentz indexes, which leads  to weaker current-meson coupling, $\lambda_{Z(1^{+-})}M_{Z(1^{+-})}< \lambda_{Z(2^{++})}$. If we study the strong decays of the $1^{+-}$ and $2^{++}$ tetraquark states with the three-point QCD sum rules by taking the pole residues $\lambda_{Z(1^{+-})}$ and $\lambda_{Z(2^{++})}$ as input parameters, the difference in the pole residues at the hadronic side is expected to be compensated by the difference in the  spectral densities at the QCD side, so the difference in the pole residues will not influence  the decay widths significantly.

The BESIII collaboration  observed the  $Z^{\pm}_c(4025)$ and $Z^{\pm}_c(4020)$ in the following processes \cite{BES1308,BES1309},
\begin{eqnarray}
e^+ e^- &\to&Z^{\pm}_c(4025)\pi^{\mp} \to (D^*\bar{D}^*)^{\pm}(0^{++},1^{+-},2^{++},0^{-+},1^{--},2^{-+},3^{--})\,\pi^{\mp} \, , \nonumber\\
e^+ e^- &\to&Z^{\pm}_c(4020)\pi^{\mp} \to (h_c\pi)^{\pm}(1^{--},0^{++},1^{+-},2^{++})\,\pi^{\mp} \, ,
\end{eqnarray}
where we present the possible quantum numbers $J^{PC}$ of the  $(D^*\bar{D}^*)^{\pm}$ and $(h_c\pi)^{\pm}$ systems in the brackets.
If the $Z^{\pm}_c(4025)$ and $Z^{\pm}_c(4020)$ are the same particle, the quantum numbers are $J^{PC}=1^{--}$, $0^{++}$, $1^{+-}$, $2^{++}$. On the other hand, the
$Z^{\pm}_c(4025)\pi^{\mp}$ and $Z^{\pm}_c(4020)\pi^{\mp}$ systems have the quantum numbers $J^{PC}=1^{--}$,  then the  survived quantum numbers of  the $Z^{\pm}_c(4025)$ and $Z^{\pm}_c(4020)$ are $J^{PC}=1^{--}$, $1^{+-}$ and $2^{++}$. The predictions  based on the QCD sum rules reduce the possible quantum numbers of the  $Z_c(4025)$ and $Z_c(4020)$  to $J^{PC}=1^{+-}$ and $2^{++}$ \cite{Wang1311}.
The strong decays
\begin{eqnarray}
Y(4260)/\gamma^*(4260) &\to& Z_c^{\pm}(4025/4020)(2^{++})\,\pi^{\mp} \, ,
\end{eqnarray}
 take place through relative D-wave, and are  kinematically suppressed in the phase-space compared to the strong  decays
 \begin{eqnarray}
Y(4260)/\gamma^*(4260) &\to& Z_c^{\pm}(4025/4020)(1^{+-})\,\pi^{\mp} \, .
\end{eqnarray}
  The $2^{++}$ assignment is disfavored, but not excluded.

In the following, we list out the possible strong decays of the $Z^{\pm}_c(4025)$ and $Z^{\pm}_c(3900)$ in the case of the $J^{PC}=1^{+-}$ assignment.
\begin{eqnarray}
Z^{\pm}_c(4025)(1^{+-}) &\to& h_c({\rm 1P})\pi^{\pm}\, , \, J/\psi\pi^{\pm}\, , \, \eta_c \rho^{\pm}\, , \, \eta_c(\pi\pi)_{\rm P}^{\pm}
\, , \,\chi_{c1}(\pi\pi)_{\rm P}^{\pm}\, , \,(D\bar{D}^*)^{\pm}\, , \,(D^*\bar{D}^*)^{\pm}\, , \nonumber\\
Z^{\pm}_c(3900)(1^{+-}) &\to& h_c({\rm 1P})\pi^{\pm}\, , \, J/\psi\pi^{\pm}\, , \, \eta_c \rho^{\pm} \, , \, \eta_c(\pi\pi)_{\rm P}^{\pm} \, , \,\chi_{c1}(\pi\pi)_{\rm P}^{\pm}\, ,
\end{eqnarray}
where the  $(\pi\pi)_{\rm P}$ denotes  the P-wave $\pi\pi$ systems have the same quantum numbers of the $\rho$.  We take  the $Z_c(4025)$ and $Z_c(4020)$ as the same particle in the $J^{PC}=1^{+-}$ assignment, and will denote them as $Z_c(4025)$.
 In Ref.\cite{WangHuangTao}, we observe that the $Z_c(3900)$ couples  to the axial-vector current $\eta_{1^{+-}}^{\mu}$.
In the following, we  perform Fierz re-arrangement  both in the color and Dirac-spinor  spaces to obtain the  result,
\begin{eqnarray}
\eta_{1^{+-}}^{\mu}&=&\frac{\epsilon^{ijk}\epsilon^{imn}}{\sqrt{2}}\left\{u^jC\gamma_5 c^k \bar{d}^m\gamma^\mu C \bar{c}^n-u^jC\gamma^\mu c^k\bar{d}^m\gamma_5 C \bar{c}^n \right\} \, , \nonumber\\
 &=&\frac{1}{2\sqrt{2}}\left\{\,i\bar{c}i\gamma_5 c\,\bar{d}\gamma^\mu u-i\bar{c} \gamma^\mu c\,\bar{d}i\gamma_5 u+\bar{c} u\,\bar{d}\gamma^\mu\gamma_5 c-\bar{c} \gamma^\mu \gamma_5u\,\bar{d}c\right. \nonumber\\
&&\left. - i\bar{c}\gamma_\nu\gamma_5c\, \bar{d}\sigma^{\mu\nu}u+i\bar{c}\sigma^{\mu\nu}c\, \bar{d}\gamma_\nu\gamma_5u
- i \bar{c}\sigma^{\mu\nu}\gamma_5u\,\bar{d}\gamma_\nu c+i\bar{c}\gamma_\nu u\, \bar{d}\sigma^{\mu\nu}\gamma_5c   \,\right\} \, ,
\end{eqnarray}
the components $\bar{c}\sigma^{\mu\nu}c\, \bar{d}\gamma_\nu\gamma_5u$, $\bar{c} \gamma^\mu c\,\bar{d}i\gamma_5 u$, $\bar{c}i\gamma_5 c\,\bar{d}\gamma^\mu u$, $\bar{c}i\gamma_5 c\,\bar{d}\gamma^\mu u$ couple  to the  $h_c({\rm 1P})\pi^{+}$, $J/\psi\pi^{+}$, $\eta_c \rho^{+}$, $\eta_c(\pi\pi)_{\rm P}^{+}$, respectively. The strong decays
\begin{eqnarray}
Z^{\pm}_c(3900)(1^{+-}) &\to& h_c({\rm 1P})\pi^{\pm}\, , \, J/\psi\pi^{\pm}\, , \, \eta_c \rho^{\pm} \, , \, \eta_c(\pi\pi)_{\rm P}^{\pm} \, ,
\end{eqnarray}
are Okubo-Zweig-Iizuka  super-allowed, we take the decays to the $(\pi\pi)_{\rm P}^{\pm}$   final states as Okubo-Zweig-Iizuka super-allowed according to the decays $\rho \to \pi\pi$.
The BESIII collaboration observed no evidence of the $Z_c(3900)$ in the process $e^+e^- \to \pi^+\pi^- h_c$ at center-of-mass energies $(3.90-4.42)\,\rm{GeV}$  \cite{BES1309}. We expect to observe the $Z_c^{\pm}(3900)$ in the $h_c({\rm 1P})\pi^{\pm}$ final states in the futures   when a large amount of events are accumulated.
The components $\bar{c}\sigma^{\mu\nu}\gamma_5u\,\bar{d}\gamma_\nu c$ and $\bar{c}\gamma_\nu u\,\bar{d} \sigma^{\mu\nu}\gamma_5c$ couples   to the scattering state $(D^*\bar{D}^*)^+$. In the nonrelativistic and heavy quark limit, the current $\bar{c}\sigma^{\mu\nu}\gamma_5u\,\bar{d}\gamma_\nu c$ is reduced to the following form,
\begin{eqnarray}
\bar{c}\sigma^{0j}\gamma_5 u \, \bar{d}\gamma_j c&\propto&  \xi^{\dagger}_c\sigma^j\zeta_u\, \chi^{\dagger}_d\vec{\sigma}\cdot \vec{k}_{d}\sigma^j\xi_c\,\,\,\propto\,\,\, \xi^{\dagger}_c\frac{\sigma^j}{2}\zeta_u\, \chi^{\dagger}_d\frac{\sigma^j}{2}\xi_c=\vec{S}_{\bar{D}^*} \cdot \vec{S}_{D^*}  \, , \nonumber\\
\bar{c}\sigma^{ij}\gamma_5 u \, \bar{d}\gamma_j c&\propto& \epsilon^{ijk} \xi^{\dagger}_c\sigma^k\vec{\sigma}\cdot \vec{k}_{u}\zeta_u\, \chi^{\dagger}_d\vec{\sigma}\cdot \vec{k}_{d}\sigma^j\xi_c\,\,\, \propto\,\,\, \epsilon^{ijk} \xi^{\dagger}_c\frac{\sigma^k}{2}\zeta_u\, \chi^{\dagger}_d\frac{ \sigma^j}{2}\xi_c=\vec{S}_{D^*}\times \vec{S}_{\bar{D}^*} \, ,
\end{eqnarray}
where the $\xi$, $\zeta$, $\chi$ are the two-component spinors of the  quark fields, the $\vec{k}$  are the three-vectors of the  quark fields,   the $\sigma^i$ are the pauli matrixes, and the $\vec{S}$ are the spin operators. It is obvious that the currents $\bar{c}\sigma^{\mu\nu}\gamma_5u\,\bar{d}\gamma_\nu c$ and $\bar{c}\gamma_\nu u\,\bar{d}\sigma^{\mu\nu}\gamma_5 c$
couple   to the $J^P=0^+$ and $1^+$ $(D^*\bar{D}^*)^+$ states. However, the strong decays $Z^{\pm}_c(3900)(1^{+-}) \to(D^*\bar{D}^*)^{\pm}$ are kinematically forbidden. The $Z_c(4025)$  and $Z_c(3900)$ have the same quantum numbers and analogous  strong decays but different masses and quark configurations.

We can search for the $Z^{\pm}_c(4025)(1^{+-})$ in the final states
$ h_c({\rm 1P})\pi^{\pm}$, $\chi_{c1}\pi^{\pm}$, $J/\psi\pi^{\pm}$, $\eta_c \rho^{\pm}$, $\eta_c(\pi\pi)_{\rm P}^{\pm}$, $\chi_{c1}(\pi\pi)_{\rm P}^{\pm}$.
Now we perform Fierz re-arrangement both in the color and Dirac-spinor  spaces to obtain the following result,
\begin{eqnarray}
\eta_{1^{+-}}^{\mu\nu}&=&\frac{\epsilon^{ijk}\epsilon^{imn}}{\sqrt{2}}\left\{u^jC\gamma^\mu c^k \bar{d}^m\gamma^\nu C \bar{c}^n-u^jC\gamma^\nu c^k\bar{d}^m\gamma^\mu C \bar{c}^n \right\} \, , \nonumber\\
 &=&\frac{1}{2\sqrt{2}}\left\{\,i\bar{d}u\, \bar{c}\sigma^{\mu\nu}c +i\bar{d}\sigma^{\mu\nu}u \,\bar{c}c+i\bar{d}c\, \bar{c}\sigma^{\mu\nu}u +i\bar{d}\sigma^{\mu\nu}c \,\bar{c}u \right. \nonumber\\
 &&-\bar{c}\sigma^{\mu\nu}\gamma_5c\,\bar{d}i\gamma_5u-\bar{c}i\gamma_5 c\,\bar{d}\sigma^{\mu\nu}\gamma_5u -\bar{c}\sigma^{\mu\nu}\gamma_5u\,\bar{d}i\gamma_5c-\bar{d}i\gamma_5 c\,\bar{c}\sigma^{\mu\nu}\gamma_5u\nonumber\\
 &&+i\epsilon^{\mu\nu\alpha\beta}\bar{c}\gamma^\alpha\gamma_5c\, \bar{d}\gamma^\beta u-i\epsilon^{\mu\nu\alpha\beta}\bar{c}\gamma^\alpha c\, \bar{d}\gamma^\beta \gamma_5u\nonumber\\
 &&\left.+i\epsilon^{\mu\nu\alpha\beta}\bar{c}\gamma^\alpha\gamma_5u\, \bar{d}\gamma^\beta c-i\epsilon^{\mu\nu\alpha\beta}\bar{c}\gamma^\alpha u\, \bar{d}\gamma^\beta \gamma_5c \,\right\} \, .
\end{eqnarray}
 The scattering states $J/\psi\pi^{+}$, $\eta_c \rho^{+}$, $\eta_c(\pi\pi)_{\rm P}^{+}$,
$\chi_{c1}(\pi\pi)_{\rm P}^{+}$, $(DD^*)^+$ couple   to the components   $\bar{c}\sigma^{\mu\nu}\gamma_5c\,\bar{d}i\gamma_5u$,
$\bar{c}i\gamma_5 c\,\bar{d}\sigma^{\mu\nu}\gamma_5u $, $\bar{c}i\gamma_5 c\,\bar{d}\sigma^{\mu\nu}\gamma_5u $, $\epsilon^{\mu\nu\alpha\beta}\bar{c}\gamma^\alpha\gamma_5c\, \bar{d}\gamma^\beta u$, $\bar{c}\sigma^{\mu\nu}\gamma_5u\,\bar{d}i\gamma_5c$, respectively.
The strong decays
\begin{eqnarray}
Z^{\pm}_c(4025)(1^{+-}) &\to&  J/\psi\pi^{\pm}\, , \, \eta_c \rho^{\pm}\, , \, \eta_c(\pi\pi)_{\rm P}^{\pm}\, , \, \chi_{c1}(\pi\pi)_{\rm P}^{\pm} \, , \, (DD^*)^\pm \, ,
\end{eqnarray}
are Okubo-Zweig-Iizuka super-allowed. In this article, we take the decays to the $(\pi\pi)_{\rm P}^{\pm}/(\pi\pi\pi)_{\rm P}^{0}$   final states as Okubo-Zweig-Iizuka super-allowed according to the decays $\rho \to \pi\pi/\omega \to \pi\pi\pi$.

We can also search for the neutral partner $Z^{0}_c(4025)(1^{+-})$
in the following strong and electromagnetic decays,
\begin{eqnarray}
Z^{0}_c(4025)(1^{+-}) &\to& h_c({\rm 1P})\pi^{0} \, , \, J/\psi\pi^{0}\, , \, J/\psi \eta\, , \,\eta_c \rho^{0}\, , \, \eta_c \omega\, , \, \eta_c(\pi\pi)_{\rm P}^{0} \, , \, \chi_{cj}(\pi\pi)_{\rm P}^{0} \, , \,  \nonumber\\
&&\eta_c(\pi\pi\pi)_{\rm P}^{0} \, , \, \chi_{cj}(\pi\pi\pi)_{\rm P}^{0}\, , \,\eta_c \gamma \, , \, \chi_{cj} \gamma\, , \, (DD^*)^0\, ,
\end{eqnarray}
where the $(\pi\pi\pi)_{\rm P}$ denotes the P-wave $\pi\pi\pi$ systems with the same quantum numbers of the $\omega$.

On the other hand, if the $Z_c(4025)$ and $Z_c(4020)$ are different particles, we can search for the $Z^{\pm}_c(4025/4020)(0^{++})$ and $Z^{\pm}_c(4025/4020)(2^{++})$ in the following strong decays,
\begin{eqnarray}
Z^{\pm}_c(4025/4020)(0^{++}) &\to& \eta_c\pi^{\pm}\, , \, J/\psi\rho^{\pm}\, , \,  J/\psi(\pi\pi)_{\rm P}^{\pm}\, , \,\chi_{c1}\pi^{\pm}
\, ,\, D\bar{D}\, ,\, D^*\bar{D}^*\, , \nonumber\\
Z^{\pm}_c(4025/4020)(2^{++}) &\to& \eta_c\pi^{\pm}\, , \, J/\psi\rho^{\pm}\, , \, J/\psi(\pi\pi)_{\rm P}^{\pm}\, , \,\chi_{c1}\pi^{\pm}\, ,\, D\bar{D}\, ,\, D^*\bar{D}^*\, . \nonumber\\
\end{eqnarray}
The strong decays
\begin{eqnarray}
Y(4260)/\gamma^*(4260) &\to& Z_c^{\pm}(4025/4020)(0^{++})\,\pi^{\mp}\, ,
\end{eqnarray}
cannot take place. The $0^{++}$ assignment is excluded. If the $Z^{\pm}_c(4025/4020)(0^{++})$ states are observed one day, it is odd indeed.

In the following, we perform Fierz re-arrangement  to the tensor and scalar currents $\eta_{2^{++}}^{\mu\nu}$ and $\eta_{0^{++}}$ both in the color and Dirac-spinor  spaces to   obtain the results,
\begin{eqnarray}
\eta_{2^{++}}^{\mu\nu}&=&\frac{\epsilon^{ijk}\epsilon^{imn}}{\sqrt{2}}\left\{u^jC\gamma^\mu c^k \bar{d}^m\gamma^\nu C \bar{c}^n+u^jC\gamma^\nu c^k\bar{d}^m\gamma^\mu C \bar{c}^n \right\} \, , \nonumber\\
 &=&\frac{1}{2\sqrt{2}}\left\{\, \bar{c}\gamma^\mu\gamma_5c\, \bar{d}\gamma^\nu\gamma_5u+\bar{c}\gamma^\nu\gamma_5c\, \bar{d}\gamma^\mu\gamma_5u -\bar{c}\gamma^\mu c\, \bar{d}\gamma^\nu u-\bar{c}\gamma^\nu c\, \bar{d}\gamma^\mu u \right. \nonumber\\
 &&+\bar{c}\gamma^\mu\gamma_5u\, \bar{d}\gamma^\nu\gamma_5c+\bar{c}\gamma^\nu\gamma_5u\, \bar{d}\gamma^\mu\gamma_5c -\bar{c}\gamma^\mu u\, \bar{d}\gamma^\nu c-\bar{c}\gamma^\nu u\, \bar{d}\gamma^\mu c  \nonumber\\
 &&+g_{\alpha\beta}\left(\bar{c}\sigma^{\mu\alpha}c\, \bar{d}\sigma^{\nu\beta}u+\bar{c}\sigma^{\nu\alpha}c\, \bar{d}\sigma^{\mu\beta}u+\bar{c}\sigma^{\mu\alpha}u\, \bar{d}\sigma^{\nu\beta}c+\bar{c}\sigma^{\nu\alpha}u\, \bar{d}\sigma^{\mu\beta}c\right) \nonumber\\
 &&+g^{\mu\nu}\left( \bar{c}c\,\bar{d}u+\bar{c}i\gamma_5c\,\bar{d}i\gamma_5u+\bar{c}\gamma_{\alpha} c\,\bar{d}\gamma^{\alpha}u-\bar{c}\gamma_{\alpha}\gamma_5 c\,\bar{d}\gamma^{\alpha}\gamma_5u-\frac{1}{2}\bar{c}\sigma_{\alpha\beta} c\,\bar{d}\sigma^{\alpha\beta}u\right. \nonumber\\
 &&\left.\left.+\bar{c}u\,\bar{d}c+\bar{c}i\gamma_5u\,\bar{d}i\gamma_5c+\bar{c}\gamma_{\alpha} u\,\bar{d}\gamma^{\alpha}c-\bar{c}\gamma_{\alpha}\gamma_5 u\,\bar{d}\gamma^{\alpha}\gamma_5c-\frac{1}{2}\bar{c}\sigma_{\alpha\beta} u\,\bar{d}\sigma^{\alpha\beta}c\right) \right\} \, ,
\end{eqnarray}
\begin{eqnarray}
\eta_{0^{++}}&=&u^jC\gamma_\mu c^k \bar{d}^m\gamma^\mu C \bar{c}^n \, , \nonumber\\
 &=&  \bar{c}c\,\bar{d}u+\bar{c}i\gamma_5c\,\bar{d}i\gamma_5u+\frac{1}{2}\bar{c}\gamma_{\alpha} c\,\bar{d}\gamma^{\alpha}u-\frac{1}{2}\bar{c}\gamma_{\alpha}\gamma_5 c\,\bar{d}\gamma^{\alpha}\gamma_5u \nonumber\\
 &&+\bar{c}u\,\bar{d}c+\bar{c}i\gamma_5u\,\bar{d}i\gamma_5c+\frac{1}{2}\bar{c}\gamma_{\alpha} u\,\bar{d}\gamma^{\alpha}c-\frac{1}{2}\bar{c}\gamma_{\alpha}\gamma_5 u\,\bar{d}\gamma^{\alpha}\gamma_5c \, ,
\end{eqnarray}
where  we add  the quantum numbers $2^{++}$ and $0^{++}$ as  subscripts    to show the $J^{PC}$ explicitly. The currents $\eta_{2^{++}}^{\mu\nu}$ and $\eta_{0^{++}}$ couple  to the
$Z^{\pm}_c(4025/4020)(2^{++})$ and $Z^{\pm}_c(3850)(0^{++})$, respectively. We denote the scalar hidden charmed tetraquark states with the mass $3850\,\rm{MeV}$ as the $Z_c(3850)$, see Table 1.
Then we obtain the Okubo-Zweig-Iizuka super-allowed decays by taking into account the couplings to the meson-meson pairs,
\begin{eqnarray}
Z^{\pm}_c(4025/4020)(2^{++}) &\to& \eta_c\pi^{\pm}\, , \, J/\psi\rho^{\pm}\, , \, J/\psi(\pi\pi)_{\rm P}^{\pm}\, , \,\chi_{c1}\pi^{\pm}\, ,\, D\bar{D}\, ,\, D^*\bar{D}^*\, ,\nonumber\\
Z^{\pm}_c(3850)(0^{++}) &\to& \eta_c\pi^{\pm}\, , \,  J/\psi(\pi\pi)_{\rm P}^{\pm}\, , \,\chi_{c1}\pi^{\pm}
\, ,\, D\bar{D}\, .
\end{eqnarray}
 We can search for the scalar and tensor tetraquark states in the futures at the BESIII, LHCb and Belle-II.
 The diquark-antidiquark type current with special quantum numbers couples    to a special tetraquark state, while the current can be re-arranged both in the color and Dirac-spinor  spaces, and changed  to a current as a special superposition of   color  singlet-singlet type currents.   The color  singlet-singlet type currents couple to the meson-meson pairs. The
diquark-antidiquark type tetraquark state can be taken as a special superposition of a series of  meson-meson pairs, and embodies  the net effects. The decays to its components (meson-meson pairs) are Okubo-Zweig-Iizuka super-allowed,  the kinematically allowed decays take place easily.

The ground state masses of the scalar tetraquark states are $M_{J=0}=3832\,\rm{MeV}$ or $3723\,\rm{MeV}$ in the constituent diquark model  \cite{Maiani-3872}, $M_{J=0}=3852\,\rm{MeV}$ in the relativistic quark model \cite{Ebert-3872},  $M_{J=0}=3729\,\rm{MeV}$ in the  relativized  quark model \cite{Carlucci-J0}. The present prediction $M_{J=0}=\left(3.85^{+0.15}_{-0.09}\right)\,\rm{GeV}$ is compatible with the values from Refs.\cite{Maiani-3872,Ebert-3872}.

\section{Conclusion}
In this article, we distinguish the charge conjugations of the interpolating currents,  calculate the contributions of the vacuum condensates up to dimension-10  in the operator product expansion,  study the   $C\gamma_\mu-C\gamma_\nu$ type scalar, axial-vector and tensor tetraquark states in details with the QCD sum rules. In calculations,  we use the  formula $\mu=\sqrt{M^2_{X/Y/Z}-(2{\mathbb{M}}_c)^2}$ suggested in our previous work to determine  the energy scales of the QCD spectral densities,  the $\mu$ can be interpreted  as the virtuality (or bound energy not as robust) in the heavy quark limit.
The predictions  $M_{J=2} =\left(4.02^{+0.09}_{-0.09}\right)\,\rm{GeV}$, $M_{J=1} =\left(4.02^{+0.07}_{-0.08}\right)\,\rm{GeV}$ are  consistent with the experimental values $M_{Z_c(4025)}=(4026.3\pm2.6\pm3.7)\,\rm{MeV}$, $M_{Z_c(4020)}=(4022.9\pm 0.8\pm 2.7)\,\rm{MeV}$ from the BESIII collaboration, which favor assigning  the $Z_c(4020)$ and $Z_c(4025)$   as the $J^{PC}=1^{+-}$ or $2^{++}$   diquark-antidiquark type tetraquark states.
 The prediction $M_{J=0}=\left(3.85^{+0.15}_{-0.09}\right)\,\rm{GeV}$ disfavors assigning  the  $Z(4050)$ and $Z(4250)$ as the $J^{PC}=0^{++}$ diquark-antidiquark type tetraquark states.  There is no candidate for the scalar hidden charmed tetraquark state, the
  prediction $M_{J=0}=\left(3.85^{+0.15}_{-0.09}\right)\,\rm{GeV}$ can be confronted with the experimental data in the futures at the BESIII, LHCb and Belle-II.
  More experimental data on the spin and parity are stilled needed to identify the $Z_c(4020)$ and $Z_c(4025)$, while the $Z(4050)$ and $Z(4250)$ still
 need  confirmation. 
 Furthermore, we discuss the strong decays of the $0^{++}$, $1^{+-}$, $2^{++}$ diquark-antidiquark type tetraquark states in details, which are of phenomenological interest. 
  The  pole residues can be taken as   basic input parameters to study relevant processes of the scalar, axial-vector and tensor tetraquark states with the three-point QCD sum rules.

\section*{Appendix}
The spectral densities at the level of the quark-gluon degrees of
freedom,

\begin{eqnarray}
\rho^{J=2}_{0}(s)&=&\frac{1}{15360\pi^6}\int_{y_i}^{y_f}dy \int_{z_i}^{1-y}dz \, yz\, (1-y-z)^3\left(s-\overline{m}_c^2\right)^2\left(293s^2-190s\overline{m}_c^2+17\overline{m}_c^4 \right)  \nonumber\\
&&+\frac{1}{5120\pi^6} \int_{y_i}^{y_f}dy \int_{z_i}^{1-y}dz \, yz \,(1-y-z)^2\left(s-\overline{m}_c^2\right)^4    \, ,
\end{eqnarray}

\begin{eqnarray}
\rho_{3}^{J=2}(s)&=&-\frac{m_c\langle \bar{q}q\rangle}{16\pi^4}\int_{y_i}^{y_f}dy \int_{z_i}^{1-y}dz \, (y+z)(1-y-z)\left(s-\overline{m}_c^2\right)\left(3s-\overline{m}_c^2\right)  \, ,
\end{eqnarray}

\begin{eqnarray}
\rho_{4}^{J=2}(s)&=&-\frac{m_c^2}{11520\pi^4} \langle\frac{\alpha_s GG}{\pi}\rangle\int_{y_i}^{y_f}dy \int_{z_i}^{1-y}dz \left( \frac{z}{y^2}+\frac{y}{z^2}\right)(1-y-z)^3 \nonumber\\
&&\left\{ 56s-17\overline{m}_c^2+10\overline{m}_c^4\delta\left(s-\overline{m}_c^2\right)\right\} \nonumber\\
&&-\frac{m_c^2}{3840\pi^4}\langle\frac{\alpha_s GG}{\pi}\rangle\int_{y_i}^{y_f}dy \int_{z_i}^{1-y}dz \left(\frac{z}{y^2}+\frac{y}{z^2} \right) (1-y-z)^2 \left(s-\overline{m}_c^2\right) \nonumber\\
&&-\frac{1}{15360\pi^4} \langle\frac{\alpha_s GG}{\pi}\rangle\int_{y_i}^{y_f}dy \int_{z_i}^{1-y}dz \left( y+z\right)(1-y-z)^2 \left( 185s^2-208s\overline{m}_c^2+43\overline{m}_c^4\right) \nonumber\\
&&+\frac{1}{7680\pi^4} \langle\frac{\alpha_s GG}{\pi}\rangle\int_{y_i}^{y_f}dy \int_{z_i}^{1-y}dz \left( y+z\right)(1-y-z) \left( s-\overline{m}_c^2\right)^2 \nonumber\\
&&-\frac{1}{2304\pi^4} \langle\frac{\alpha_s GG}{\pi}\rangle\int_{y_i}^{y_f}dy \int_{z_i}^{1-y}dz \left( y+z\right)(1-y-z)^2 \left( 15s^2-16s\overline{m}_c^2+3\overline{m}_c^4\right) \nonumber\\
&&-\frac{1}{13824\pi^4} \langle\frac{\alpha_s GG}{\pi}\rangle\int_{y_i}^{y_f}dy \int_{z_i}^{1-y}dz \, (1-y-z)^3 \left( 25s^2-24s\overline{m}_c^2+3\overline{m}_c^4\right)  \nonumber\\
&&-\frac{1}{6912\pi^4} \langle\frac{\alpha_s GG}{\pi}\rangle\int_{y_i}^{y_f}dy \int_{z_i}^{1-y}dz  \, yz\,(1-y-z) \left( 25s^2-24s\overline{m}_c^2+3\overline{m}_c^4\right) \nonumber\\
&&-\frac{1}{4608\pi^4} \langle\frac{\alpha_s GG}{\pi}\rangle\int_{y_i}^{y_f}dy \int_{z_i}^{1-y}dz \, (1-y-z)^2 \left( s-\overline{m}_c^2\right)^2 \nonumber\\
&&-\frac{1}{6912\pi^4} \langle\frac{\alpha_s GG}{\pi}\rangle\int_{y_i}^{y_f}dy \int_{z_i}^{1-y}dz \, yz \left( s-\overline{m}_c^2\right)\left( 13s-5\overline{m}_c^2\right) \, ,
\end{eqnarray}

\begin{eqnarray}
\rho^{J=2}_{5}(s)&=&\frac{m_c\langle \bar{q}g_s\sigma Gq\rangle}{32\pi^4}\int_{y_i}^{y_f}dy \int_{z_i}^{1-y}dz  \, (y+z) \left(2s-\overline{m}_c^2 \right) \nonumber\\
&&+\frac{m_c\langle \bar{q}g_s\sigma Gq\rangle}{144\pi^4}\int_{y_i}^{y_f}dy \int_{z_i}^{1-y}dz  \,  (1-y-z) \left(2s-\overline{m}_c^2 \right)     \, ,
\end{eqnarray}

\begin{eqnarray}
\rho_{6}^{J=2}(s)&=&\frac{m_c^2\langle\bar{q}q\rangle^2}{6\pi^2}\int_{y_i}^{y_f}dy   +\frac{g_s^2\langle\bar{q}q\rangle^2}{3240\pi^4}\int_{y_i}^{y_f}dy \int_{z_i}^{1-y}dz\, yz \left\{56s-17\overline{m}_c^2 +10\overline{m}_c^4\delta\left(s-\overline{m}_c^2 \right)\right\}\nonumber\\
&&+\frac{g_s^2\langle\bar{q}q\rangle^2}{3240\pi^4}\int_{y_i}^{y_f}dy \,y(1-y)\left(s-\widetilde{m}_c^2 \right)  \nonumber\\
&&-\frac{g_s^2\langle\bar{q}q\rangle^2}{9720\pi^4}\int_{y_i}^{y_f}dy \int_{z_i}^{1-y}dz \, (1-y-z)\left\{ 45\left(\frac{z}{y}+\frac{y}{z} \right)\left(2s-\overline{m}_c^2 \right)+\left(\frac{z}{y^2}+\frac{y}{z^2} \right)\right.\nonumber\\
&&\left.m_c^2\left[ 19+20\overline{m}_c^2\delta\left(s-\overline{m}_c^2 \right)\right]+(y+z)\left[18\left(3s-\overline{m}_c^2\right)+10\overline{m}_c^4\delta\left(s-\overline{m}_c^2\right) \right] \right\} \nonumber\\
&&-\frac{g_s^2\langle\bar{q}q\rangle^2}{9720\pi^4}\int_{y_i}^{y_f}dy \int_{z_i}^{1-y}dz \, (1-y-z)\left\{  15\left(\frac{z}{y}+\frac{y}{z} \right)\left(2s-\overline{m}_c^2 \right)+\left(\frac{z}{y^2}+\frac{y}{z^2} \right)\right. \nonumber\\
&&\left.m_c^2\left[ 6+5\overline{m}_c^2\delta\left(s-\overline{m}_c^2\right)\right]+(y+z)\left[56s-17\overline{m}_c^2 +10\overline{m}_c^4\delta\left(s-\overline{m}_c^2\right)\right] \right\}\, ,
\end{eqnarray}

\begin{eqnarray}
\rho_7^{J=2}(s)&=&\frac{m_c^3\langle\bar{q}q\rangle}{144\pi^2 T^2}\langle\frac{\alpha_sGG}{\pi}\rangle\int_{y_i}^{y_f}dy \int_{z_i}^{1-y}dz \left(\frac{y}{z^3}+\frac{z}{y^3}+\frac{1}{y^2}+\frac{1}{z^2}\right)(1-y-z)\, \overline{m}_c^2 \, \delta\left(s-\overline{m}_c^2\right)\nonumber\\
&&-\frac{m_c\langle\bar{q}q\rangle}{48\pi^2}\langle\frac{\alpha_sGG}{\pi}\rangle\int_{y_i}^{y_f}dy \int_{z_i}^{1-y}dz \left(\frac{y}{z^2}+\frac{z}{y^2}\right)(1-y-z)  \left\{1+\overline{m}_c^2\delta\left(s-\overline{m}_c^2\right) \right\}\nonumber\\
&&+\frac{m_c\langle\bar{q}q\rangle}{48\pi^2}\langle\frac{\alpha_sGG}{\pi}\rangle\int_{y_i}^{y_f}dy \int_{z_i}^{1-y}dz\left\{1+\frac{\overline{m}_c^2}{3}\delta\left(s-\overline{m}_c^2\right) \right\} \nonumber\\
&&+\frac{m_c\langle\bar{q}q\rangle}{432\pi^2}\langle\frac{\alpha_sGG}{\pi}\rangle\int_{y_i}^{y_f}dy \int_{z_i}^{1-y}dz\left(\frac{1-y}{y}+\frac{1-z}{z}\right)
\left\{1+\overline{m}_c^2\delta\left(s-\overline{m}_c^2\right) \right\}\nonumber \\
&&-\frac{m_c\langle\bar{q}q\rangle}{288\pi^2}\langle\frac{\alpha_sGG}{\pi}\rangle\int_{y_i}^{y_f}dy \left\{1+ \widetilde{m}_c^2 \, \delta \left(s-\widetilde{m}_c^2\right) \right\}\, ,
\end{eqnarray}

\begin{eqnarray}
\rho_8^{J=2}(s)&=&-\frac{m_c^2\langle\bar{q}q\rangle\langle\bar{q}g_s\sigma Gq\rangle}{12\pi^2}\int_0^1 dy \left(1+\frac{\widetilde{m}_c^2}{T^2} \right)\delta\left(s-\widetilde{m}_c^2\right)\nonumber \\
&&-\frac{ m_c^2\langle\bar{q}q\rangle\langle\bar{q}g_s\sigma Gq\rangle}{216\pi^2}\int_{0}^{1} dy \frac{1}{y(1-y)}\delta\left(s-\widetilde{m}_c^2\right)
 \, ,
\end{eqnarray}

\begin{eqnarray}
\rho_{10}^{J=2}(s)&=&\frac{m_c^2\langle\bar{q}g_s\sigma Gq\rangle^2}{96\pi^2T^6}\int_0^1 dy \, \widetilde{m}_c^4 \, \delta \left( s-\widetilde{m}_c^2\right)
\nonumber \\
&&-\frac{m_c^4\langle\bar{q}q\rangle^2}{108T^4}\langle\frac{\alpha_sGG}{\pi}\rangle\int_0^1 dy  \left\{ \frac{1}{y^3}+\frac{1}{(1-y)^3}\right\} \delta\left( s-\widetilde{m}_c^2\right)\nonumber\\
&&+\frac{m_c^2\langle\bar{q}q\rangle^2}{36T^2}\langle\frac{\alpha_sGG}{\pi}\rangle\int_0^1 dy  \left\{ \frac{1}{y^2}+\frac{1}{(1-y)^2}\right\} \delta\left( s-\widetilde{m}_c^2\right)\nonumber\\
&&-\frac{m_c^2\langle\bar{q}q\rangle^2}{324T^2}\langle\frac{\alpha_sGG}{\pi}\rangle\int_0^1 dy   \frac{1}{y(1-y)} \delta\left( s-\widetilde{m}_c^2\right)\nonumber \\
&&+\frac{m_c^2\langle\bar{q}g_s\sigma Gq\rangle^2}{864 \pi^2T^4} \int_0^1 dy   \frac{1}{y(1-y)}  \widetilde{m}_c^2 \, \delta\left( s-\widetilde{m}_c^2\right)\nonumber\\
&&+\frac{m_c^2\langle\bar{q}g_s\sigma Gq\rangle^2}{576 \pi^2T^2} \int_0^1 dy   \frac{1}{y(1-y)}   \delta\left( s-\widetilde{m}_c^2\right)\nonumber \\
&&+\frac{m_c^2\langle\bar{q} q\rangle^2}{108 T^6}\langle\frac{\alpha_sGG}{\pi}\rangle\int_0^1 dy \, \widetilde{m}_c^4 \, \delta \left( s-\widetilde{m}_c^2\right) \, ,
\end{eqnarray}

\begin{eqnarray}
\rho^{J=1}_{0}(s)&=&\frac{1}{3072\pi^6s}\int_{y_i}^{y_f}dy \int_{z_i}^{1-y}dz \, yz\, (1-y-z)^3\left(s-\overline{m}_c^2\right)^2\left(49s^2-30s\overline{m}_c^2+\overline{m}_c^4 \right)  \nonumber\\
&&+\frac{1}{3072\pi^6s} \int_{y_i}^{y_f}dy \int_{z_i}^{1-y}dz \, yz \,(1-y-z)^2\left(s-\overline{m}_c^2\right)^3\left(3s+\overline{m}_c^2\right)    \, ,
\end{eqnarray}

\begin{eqnarray}
\rho_{3}^{J=1}(s)&=&-\frac{m_c\langle \bar{q}q\rangle}{16\pi^4}\int_{y_i}^{y_f}dy \int_{z_i}^{1-y}dz \, (y+z)(1-y-z)\left(s-\overline{m}_c^2\right)  \, ,
\end{eqnarray}

\begin{eqnarray}
\rho_{4}^{J=1}(s)&=&-\frac{m_c^2}{2304\pi^4s} \langle\frac{\alpha_s GG}{\pi}\rangle\int_{y_i}^{y_f}dy \int_{z_i}^{1-y}dz \left( \frac{z}{y^2}+\frac{y}{z^2}\right)(1-y-z)^3 \nonumber\\
&&\left\{ 8s-\overline{m}_c^2+\frac{5\overline{m}_c^4}{3}\delta\left(s-\overline{m}_c^2\right)\right\} \nonumber\\
&&-\frac{m_c^2}{2304\pi^4s}\langle\frac{\alpha_s GG}{\pi}\rangle\int_{y_i}^{y_f}dy \int_{z_i}^{1-y}dz \left(\frac{z}{y^2}+\frac{y}{z^2} \right) (1-y-z)^2 \, \overline{m}_c^2 \nonumber\\
&&-\frac{1}{9216\pi^4s} \langle\frac{\alpha_s GG}{\pi}\rangle\int_{y_i}^{y_f}dy \int_{z_i}^{1-y}dz \left( y+z\right)(1-y-z)^2 \left( 5s^2-3\overline{m}_c^4\right) \nonumber\\
&&+\frac{1}{4608\pi^4s} \langle\frac{\alpha_s GG}{\pi}\rangle\int_{y_i}^{y_f}dy \int_{z_i}^{1-y}dz \left( y+z\right)(1-y-z) \left( s^2-\overline{m}_c^4\right) \nonumber\\
&&+\frac{1}{2304\pi^4} \langle\frac{\alpha_s GG}{\pi}\rangle\int_{y_i}^{y_f}dy \int_{z_i}^{1-y}dz \left( y+z\right)(1-y-z)^2 \left( 5s-4\overline{m}_c^2\right) \nonumber\\
&&+\frac{1}{41472\pi^4s} \langle\frac{\alpha_s GG}{\pi}\rangle\int_{y_i}^{y_f}dy \int_{z_i}^{1-y}dz \, (1-y-z)^3 \left( 55s^2-48s\overline{m}_c^2+3\overline{m}_c^4\right)  \nonumber\\
&&+\frac{1}{6912\pi^4s} \langle\frac{\alpha_s GG}{\pi}\rangle\int_{y_i}^{y_f}dy \int_{z_i}^{1-y}dz  \, yz\,(1-y-z) \left( 5s^2-3\overline{m}_c^4\right) \nonumber\\
&&-\frac{1}{3456\pi^4s} \langle\frac{\alpha_s GG}{\pi}\rangle\int_{y_i}^{y_f}dy \int_{z_i}^{1-y}dz \, (1-y-z)^2 \left( s-\overline{m}_c^2\right) \left( 2s-\overline{m}_c^2\right)  \nonumber\\
&&+\frac{1}{1728\pi^4s} \langle\frac{\alpha_s GG}{\pi}\rangle\int_{y_i}^{y_f}dy \int_{z_i}^{1-y}dz \, yz \left( s-\overline{m}_c^2\right)\left( 2s-\overline{m}_c^2\right) \, ,
\end{eqnarray}

\begin{eqnarray}
\rho^{J=1}_{5}(s)&=&\frac{m_c\langle \bar{q}g_s\sigma Gq\rangle}{64\pi^4}\int_{y_i}^{y_f}dy \int_{z_i}^{1-y}dz  \, (y+z)\nonumber\\
&&
-\frac{m_c\langle \bar{q}g_s\sigma Gq\rangle}{288\pi^4}\int_{y_i}^{y_f}dy \int_{z_i}^{1-y}dz  \,  (1-y-z)      \, ,
\end{eqnarray}

\begin{eqnarray}
\rho_{6}^{J=1}(s)&=& \frac{g_s^2\langle\bar{q}q\rangle^2}{648\pi^4s}\int_{y_i}^{y_f}dy \int_{z_i}^{1-y}dz\, yz \left\{8s-\overline{m}_c^2 +\frac{5\overline{m}_c^4}{3}\delta\left(s-\overline{m}_c^2 \right)\right\}\nonumber\\
&&+\frac{g_s^2\langle\bar{q}q\rangle^2}{1944\pi^4s}\int_{y_i}^{y_f}dy \,y(1-y) \,\widetilde{m}_c^2   \nonumber\\
&&-\frac{g_s^2\langle\bar{q}q\rangle^2}{1296\pi^4}\int_{y_i}^{y_f}dy \int_{z_i}^{1-y}dz \, (1-y-z)\left\{ 3\left(\frac{z}{y}+\frac{y}{z} \right)  +\left(\frac{z}{y^2}+\frac{y}{z^2} \right)m_c^2  \delta\left(s-\overline{m}_c^2 \right)\right.\nonumber\\
&&\left. +(y+z)\left[8+2\overline{m}_c^2\delta\left(s-\overline{m}_c^2\right) \right] \right\} \nonumber\\
&&-\frac{g_s^2\langle\bar{q}q\rangle^2}{11664\pi^4s}\int_{y_i}^{y_f}dy \int_{z_i}^{1-y}dz \, (1-y-z)\left\{  27\left(\frac{z}{y}+\frac{y}{z} \right)s+11\left(\frac{z}{y^2}+\frac{y}{z^2} \right)\right. \nonumber\\
&&\left.m_c^2\overline{m}_c^2\delta\left(s-\overline{m}_c^2\right)+(y+z)\left[6\left(8s-\overline{m}_c^2\right) +10\overline{m}_c^4\delta\left(s-\overline{m}_c^2\right)\right] \right\}\, ,
\end{eqnarray}

\begin{eqnarray}
\rho_7^{J=1}(s)&=&\frac{m_c^3\langle\bar{q}q\rangle}{288\pi^2 T^2}\langle\frac{\alpha_sGG}{\pi}\rangle\int_{y_i}^{y_f}dy \int_{z_i}^{1-y}dz \left(\frac{y}{z^3}+\frac{z}{y^3}+\frac{1}{y^2}+\frac{1}{z^2}\right)(1-y-z) \delta\left(s-\overline{m}_c^2\right)\nonumber\\
&&-\frac{m_c\langle\bar{q}q\rangle}{96\pi^2}\langle\frac{\alpha_sGG}{\pi}\rangle\int_{y_i}^{y_f}dy \int_{z_i}^{1-y}dz \left(\frac{y}{z^2}+\frac{z}{y^2}\right)(1-y-z)  \delta\left(s-\overline{m}_c^2\right)\nonumber\\
&&-\frac{m_c\langle\bar{q}q\rangle}{288\pi^2}\langle\frac{\alpha_sGG}{\pi}\rangle\int_{y_i}^{y_f}dy \int_{z_i}^{1-y}dz\delta\left(s-\overline{m}_c^2\right) \nonumber\\
&&-\frac{m_c\langle\bar{q}q\rangle}{864\pi^2}\langle\frac{\alpha_sGG}{\pi}\rangle\int_{y_i}^{y_f}dy \int_{z_i}^{1-y}dz\left(\frac{1-y}{y}+\frac{1-z}{z}\right)
\delta\left(s-\overline{m}_c^2\right) \nonumber \\
&&-\frac{m_c\langle\bar{q}q\rangle}{576\pi^2}\langle\frac{\alpha_sGG}{\pi}\rangle\int_{y_i}^{y_f}dy   \delta \left(s-\widetilde{m}_c^2\right) \, ,
\end{eqnarray}

\begin{eqnarray}
\rho^{J=0}_{0}(s)&=&\frac{1}{256\pi^6}\int_{y_i}^{y_f}dy \int_{z_i}^{1-y}dz \, yz\, (1-y-z)^3\left(s-\overline{m}_c^2\right)^2\left(7s^2-6s\overline{m}_c^2+\overline{m}_c^4 \right)  \nonumber\\
&&+\frac{1}{256\pi^6} \int_{y_i}^{y_f}dy \int_{z_i}^{1-y}dz \, yz \,(1-y-z)^2\left(s-\overline{m}_c^2\right)^3 \left(3s-\overline{m}_c^2\right)   \, ,
\end{eqnarray}

\begin{eqnarray}
\rho_{3}^{J=0}(s)&=&-\frac{m_c\langle \bar{q}q\rangle}{8\pi^4}\int_{y_i}^{y_f}dy \int_{z_i}^{1-y}dz \, (y+z)(1-y-z)\left(s-\overline{m}_c^2\right)\left(2s-\overline{m}_c^2\right)  \, ,
\end{eqnarray}

\begin{eqnarray}
\rho_{4}^{J=0}(s)&=&-\frac{m_c^2}{192\pi^4} \langle\frac{\alpha_s GG}{\pi}\rangle\int_{y_i}^{y_f}dy \int_{z_i}^{1-y}dz \left( \frac{z}{y^2}+\frac{y}{z^2}\right)(1-y-z)^3 \nonumber\\
&&\left\{ 2s-\overline{m}_c^2+\frac{\overline{m}_c^4}{6}\delta\left(s-\overline{m}_c^2\right)\right\} \nonumber\\
&&-\frac{m_c^2}{384\pi^4}\langle\frac{\alpha_s GG}{\pi}\rangle\int_{y_i}^{y_f}dy \int_{z_i}^{1-y}dz \left(\frac{z}{y^2}+\frac{y}{z^2} \right) (1-y-z)^2 \left(3s-2\overline{m}_c^2\right) \nonumber\\
&&-\frac{1}{768\pi^4} \langle\frac{\alpha_s GG}{\pi}\rangle\int_{y_i}^{y_f}dy \int_{z_i}^{1-y}dz \left( y+z\right)(1-y-z)^2 \left( 10s^2-12s\overline{m}_c^2+3\overline{m}_c^4\right) \nonumber\\
&&+\frac{1}{384\pi^4} \langle\frac{\alpha_s GG}{\pi}\rangle\int_{y_i}^{y_f}dy \int_{z_i}^{1-y}dz \left( y+z\right)(1-y-z) \left( s-\overline{m}_c^2\right)\left( 2s-\overline{m}_c^2\right) \nonumber\\
&&+\frac{1}{384\pi^4} \langle\frac{\alpha_s GG}{\pi}\rangle\int_{y_i}^{y_f}dy \int_{z_i}^{1-y}dz \left( y+z\right)(1-y-z)^2 \left( 10s^2-12s\overline{m}_c^2+3\overline{m}_c^4\right) \nonumber\\
&&+\frac{1}{3456\pi^4} \langle\frac{\alpha_s GG}{\pi}\rangle\int_{y_i}^{y_f}dy \int_{z_i}^{1-y}dz \, (1-y-z)^3 \left( 10s^2-12s\overline{m}_c^2+3\overline{m}_c^4\right)  \nonumber\\
&&+\frac{1}{576\pi^4} \langle\frac{\alpha_s GG}{\pi}\rangle\int_{y_i}^{y_f}dy \int_{z_i}^{1-y}dz  \, yz\,(1-y-z) \left( 10s^2-12s\overline{m}_c^2+3\overline{m}_c^4\right) \nonumber\\
&&+\frac{1}{576\pi^4} \langle\frac{\alpha_s GG}{\pi}\rangle\int_{y_i}^{y_f}dy \int_{z_i}^{1-y}dz \, (1-y-z)^2 \left( s-\overline{m}_c^2\right)
\left( 2s-\overline{m}_c^2\right) \nonumber\\
&&+\frac{1}{288\pi^4} \langle\frac{\alpha_s GG}{\pi}\rangle\int_{y_i}^{y_f}dy \int_{z_i}^{1-y}dz \, yz \left( s-\overline{m}_c^2\right)\left( 2s-\overline{m}_c^2\right) \, ,
\end{eqnarray}

\begin{eqnarray}
\rho^{J=0}_{5}(s)&=&\frac{m_c\langle \bar{q}g_s\sigma Gq\rangle}{32\pi^4}\int_{y_i}^{y_f}dy \int_{z_i}^{1-y}dz  \, (y+z) \left(3s-2\overline{m}_c^2 \right) \nonumber\\
&&-\frac{m_c\langle \bar{q}g_s\sigma Gq\rangle}{48\pi^4}\int_{y_i}^{y_f}dy \int_{z_i}^{1-y}dz  \,  (1-y-z) \left(3s-2\overline{m}_c^2 \right)     \, ,
\end{eqnarray}

\begin{eqnarray}
\rho_{6}^{J=0}(s)&=&\frac{m_c^2\langle\bar{q}q\rangle^2}{3\pi^2}\int_{y_i}^{y_f}dy   +\frac{g_s^2\langle\bar{q}q\rangle^2}{54\pi^4}\int_{y_i}^{y_f}dy \int_{z_i}^{1-y}dz\, yz \left\{2s-\overline{m}_c^2 +\frac{\overline{m}_c^4}{6}\delta\left(s-\overline{m}_c^2 \right)\right\}\nonumber\\
&&+\frac{g_s^2\langle\bar{q}q\rangle^2}{324\pi^4}\int_{y_i}^{y_f}dy \,y(1-y)\left(3s-2\widetilde{m}_c^2 \right)  \nonumber\\
&&-\frac{g_s^2\langle\bar{q}q\rangle^2}{648\pi^4}\int_{y_i}^{y_f}dy \int_{z_i}^{1-y}dz \, (1-y-z)\left\{ 3\left(\frac{z}{y}+\frac{y}{z} \right)\left(3s-2\overline{m}_c^2 \right)+\left(\frac{z}{y^2}+\frac{y}{z^2} \right)\right.\nonumber\\
&&\left.m_c^2\left[ 2+ \overline{m}_c^2\delta\left(s-\overline{m}_c^2 \right)\right]+(y+z)\left[12\left(2s-\overline{m}_c^2\right)+2\overline{m}_c^4\delta\left(s-\overline{m}_c^2\right) \right] \right\} \nonumber\\
&&-\frac{g_s^2\langle\bar{q}q\rangle^2}{1944\pi^4}\int_{y_i}^{y_f}dy \int_{z_i}^{1-y}dz \, (1-y-z)\left\{  15\left(\frac{z}{y}+\frac{y}{z} \right)\left(3s-2\overline{m}_c^2 \right)+7\left(\frac{z}{y^2}+\frac{y}{z^2} \right)\right. \nonumber\\
&&\left.m_c^2\left[ 2+\overline{m}_c^2\delta\left(s-\overline{m}_c^2\right)\right]+(y+z)\left[12\left(2s-\overline{m}_c^2\right) +2\overline{m}_c^4\delta\left(s-\overline{m}_c^2\right)\right] \right\}\, ,
\end{eqnarray}

\begin{eqnarray}
\rho_7^{J=0}(s)&=&\frac{m_c^3\langle\bar{q}q\rangle}{144\pi^2  }\langle\frac{\alpha_sGG}{\pi}\rangle\int_{y_i}^{y_f}dy \int_{z_i}^{1-y}dz \left(\frac{y}{z^3}+\frac{z}{y^3}+\frac{1}{y^2}+\frac{1}{z^2}\right)(1-y-z)\nonumber\\
&&\left(1+\frac{ \overline{m}_c^2}{T^2}\right) \delta\left(s-\overline{m}_c^2\right)\nonumber\\
&&-\frac{m_c\langle\bar{q}q\rangle}{48\pi^2}\langle\frac{\alpha_sGG}{\pi}\rangle\int_{y_i}^{y_f}dy \int_{z_i}^{1-y}dz \left(\frac{y}{z^2}+\frac{z}{y^2}\right)(1-y-z)  \left\{2+\overline{m}_c^2\delta\left(s-\overline{m}_c^2\right) \right\}\nonumber\\
&&+\frac{m_c\langle\bar{q}q\rangle}{48\pi^2}\langle\frac{\alpha_sGG}{\pi}\rangle\int_{y_i}^{y_f}dy \int_{z_i}^{1-y}dz\left\{2+ \overline{m}_c^2 \delta\left(s-\overline{m}_c^2\right) \right\} \nonumber\\
&&-\frac{m_c\langle\bar{q}q\rangle}{144\pi^2}\langle\frac{\alpha_sGG}{\pi}\rangle\int_{y_i}^{y_f}dy \int_{z_i}^{1-y}dz\left(\frac{1-y}{y}+\frac{1-z}{z}\right)
\left\{2+\overline{m}_c^2\delta\left(s-\overline{m}_c^2\right) \right\}\nonumber \\
&&-\frac{m_c\langle\bar{q}q\rangle}{288\pi^2}\langle\frac{\alpha_sGG}{\pi}\rangle\int_{y_i}^{y_f}dy \left\{2+ \widetilde{m}_c^2 \, \delta \left(s-\widetilde{m}_c^2\right) \right\}\, ,
\end{eqnarray}

\begin{eqnarray}
\rho_8^{J=0}(s)&=&-\frac{m_c^2\langle\bar{q}q\rangle\langle\bar{q}g_s\sigma Gq\rangle}{6\pi^2}\int_0^1 dy \left(1+\frac{\widetilde{m}_c^2}{T^2} \right)\delta\left(s-\widetilde{m}_c^2\right)\nonumber \\
&&+\frac{ m_c^2\langle\bar{q}q\rangle\langle\bar{q}g_s\sigma Gq\rangle}{36\pi^2}\int_{0}^{1} dy \frac{1}{y(1-y)}\delta\left(s-\widetilde{m}_c^2\right)
 \, ,
\end{eqnarray}

\begin{eqnarray}
\rho_{10}^{J=0}(s)&=&\frac{m_c^2\langle\bar{q}g_s\sigma Gq\rangle^2}{48\pi^2T^6}\int_0^1 dy \, \widetilde{m}_c^4 \, \delta \left( s-\widetilde{m}_c^2\right)
\nonumber \\
&&-\frac{m_c^4\langle\bar{q}q\rangle^2}{54T^4}\langle\frac{\alpha_sGG}{\pi}\rangle\int_0^1 dy  \left\{ \frac{1}{y^3}+\frac{1}{(1-y)^3}\right\} \delta\left( s-\widetilde{m}_c^2\right)\nonumber\\
&&+\frac{m_c^2\langle\bar{q}q\rangle^2}{18T^2}\langle\frac{\alpha_sGG}{\pi}\rangle\int_0^1 dy  \left\{ \frac{1}{y^2}+\frac{1}{(1-y)^2}\right\} \delta\left( s-\widetilde{m}_c^2\right)\nonumber\\
&&+\frac{m_c^2\langle\bar{q}q\rangle^2}{54T^2}\langle\frac{\alpha_sGG}{\pi}\rangle\int_0^1 dy   \frac{1}{y(1-y)} \delta\left( s-\widetilde{m}_c^2\right)\nonumber \\
&&-\frac{m_c^2\langle\bar{q}g_s\sigma Gq\rangle^2}{144 \pi^2T^4} \int_0^1 dy   \frac{1}{y(1-y)}  \widetilde{m}_c^2 \, \delta\left( s-\widetilde{m}_c^2\right)\nonumber\\
&&+\frac{m_c^2\langle\bar{q}g_s\sigma Gq\rangle^2}{32 \pi^2T^2} \int_0^1 dy   \frac{1}{y(1-y)}   \delta\left( s-\widetilde{m}_c^2\right)\nonumber \\
&&+\frac{m_c^2\langle\bar{q} q\rangle^2}{54 T^6}\langle\frac{\alpha_sGG}{\pi}\rangle\int_0^1 dy \, \widetilde{m}_c^4 \, \delta \left( s-\widetilde{m}_c^2\right) \, ,
\end{eqnarray}
the subscripts  $0$, $3$, $4$, $5$, $6$, $7$, $8$, $10$ denote the dimensions of the  vacuum condensates, $y_{f}=\frac{1+\sqrt{1-4m_c^2/s}}{2}$,
$y_{i}=\frac{1-\sqrt{1-4m_c^2/s}}{2}$, $z_{i}=\frac{y
m_c^2}{y s -m_c^2}$, $\overline{m}_c^2=\frac{(y+z)m_c^2}{yz}$,
$ \widetilde{m}_c^2=\frac{m_c^2}{y(1-y)}$, $\int_{y_i}^{y_f}dy \to \int_{0}^{1}dy$, $\int_{z_i}^{1-y}dz \to \int_{0}^{1-y}dz$ when the $\delta$ functions $\delta\left(s-\overline{m}_c^2\right)$ and $\delta\left(s-\widetilde{m}_c^2\right)$ appear.
The condensates $\langle \frac{\alpha_s}{\pi}GG\rangle$, $\langle \bar{q}q\rangle\langle \frac{\alpha_s}{\pi}GG\rangle$,
$\langle \bar{q}q\rangle^2\langle \frac{\alpha_s}{\pi}GG\rangle$, $\langle \bar{q} g_s \sigma Gq\rangle^2$ and $g_s^2\langle \bar{q}q\rangle^2$ are the vacuum expectations
of the operators of the order
$\mathcal{O}(\alpha_s)$.  The four-quark condensate $g_s^2\langle \bar{q}q\rangle^2$ comes from the terms
$\langle \bar{q}\gamma_\mu t^a q g_s D_\eta G^a_{\lambda\tau}\rangle$, $\langle\bar{q}_jD^{\dagger}_{\mu}D^{\dagger}_{\nu}D^{\dagger}_{\alpha}q_i\rangle$  and
$\langle\bar{q}_jD_{\mu}D_{\nu}D_{\alpha}q_i\rangle$, rather than comes from the perturbative corrections of $\langle \bar{q}q\rangle^2$.
 The condensates $\langle g_s^3 GGG\rangle$, $\langle \frac{\alpha_s GG}{\pi}\rangle^2$,
 $\langle \frac{\alpha_s GG}{\pi}\rangle\langle \bar{q} g_s \sigma Gq\rangle$ have the dimensions 6, 8, 9 respectively,  but they are   the vacuum expectations
of the operators of the order    $\mathcal{O}( \alpha_s^{3/2})$, $\mathcal{O}(\alpha_s^2)$, $\mathcal{O}( \alpha_s^{3/2})$ respectively, and discarded.  We take
the truncations $n\leq 10$ and $k\leq 1$ in a consistent way,
the operators of the orders $\mathcal{O}( \alpha_s^{k})$ with $k> 1$ are  discarded. Furthermore,  the values of the  condensates $\langle g_s^3 GGG\rangle$, $\langle \frac{\alpha_s GG}{\pi}\rangle^2$,
 $\langle \frac{\alpha_s GG}{\pi}\rangle\langle \bar{q} g_s \sigma Gq\rangle$   are very small, and they can be  neglected safely.

\section*{Acknowledgements}
This  work is supported by National Natural Science Foundation,
Grant Numbers 11375063, and Natural Science Foundation of Hebei province, Grant Number A2014502017.
 The author would like to thank Prof. T. Huang for suggesting this subject.


\begin{thebibliography}{99}

\bibitem{Belle0806} R. Mizuk et al, Phys. Rev. {\bf D78} (2008) 072004.

\bibitem{Wang4250} Z. G. Wang, Eur. Phys. J. {\bf C59} (2009) 675;
Z. G. Wang, Eur. Phys. J. {\bf C62} (2009) 375;
Z. G. Wang, Phys. Rev. {\bf D79} (2009) 094027.


\bibitem{Wang4250-EPJC} Z. G. Wang, Eur. Phys. J. {\bf C67} (2010) 411.


\bibitem{Tetraquark4250} D. Ebert, R. N. Faustov and V. O. Galkin, Eur. Phys. J. {\bf C58} (2008) 399.


\bibitem{Molecule4250} S. H. Lee, K. Morita and M. Nielsen, Nucl. Phys. {\bf A815} (2009) 29;
S. H. Lee, K. Morita and M. Nielsen, Phys. Rev. {\bf D78} (2008) 076001;
G. J. Ding, Phys. Rev. {\bf D79} (2009) 014001;
Y. R. Liu and Z. Y. Zhang, Phys. Rev. {\bf C80} (2009) 015208;
X. Liu, Z. G. Luo, Y. R. Liu and S. L. Zhu, Eur. Phys. J. {\bf C61} (2009) 411;
T. F. Carames, A. Valcarce and J. Vijande, Phys. Rev. {\bf D82} (2010) 054032.



\bibitem{BaBar1111} J. P. Lees   et al,  Phys. Rev. {\bf D85} (2012) 052003.


\bibitem{BES1308}   M. Ablikim  et al, Phys. Rev. Lett. {\bf 112} (2014) 132001.

\bibitem{BES1309}  M. Ablikim  et al, Phys. Rev. Lett. {\bf 111} (2013) 242001.

\bibitem{Rescatter}  G. Li, Eur. Phys. J. {\bf C73} (2013) 2621;
X. Wang, Y. Sun, D. Y. Chen, X. Liu and T. Matsuki,  Eur. Phys. J. {\bf C74} (2014) 2761.

\bibitem{Molecule} F. K. Guo, C. Hidalgo-Duque, J. Nieves and M. P. Valderrama, Phys. Rev. {\bf D88} (2013) 054007;
J. He, X. Liu, Z. F. Sun and S. L. Zhu, Eur. Phys. J. {\bf C73} (2013) 2635;
C. Y. Cui, Y. L. Liu and M. Q. Huang, Eur. Phys. J. {\bf C73} (2013) 2661;
W. Chen, T. G. Steele, M. L. Du and S. L. Zhu, Eur. Phys. J. {\bf C74} (2014) 2773;
K. P. Khemchandani, A. Martinez Torres, M. Nielsen and F. S. Navarra, Phys. Rev. {\bf D89} (2014) 014029;
A. Martinez Torres, K. P. Khemchandani, F. S. Navarra, M. Nielsen and E. Oset, Phys. Rev. {\bf D89} (2014) 014025.

\bibitem{Tetraquark-Qiao} C. F. Qiao and L. Tang, Eur. Phys. J. {\bf C74} (2014) 2810.

 \bibitem{Wang1311} Z. G. Wang,  Eur. Phys. J. {\bf C74} (2014)  2874.
.

\bibitem{Swanson2006} E. S. Swanson, Phys. Rept. {\bf 429} (2006) 243;
S. Godfrey and S. L. Olsen, Ann. Rev. Nucl. Part. Sci. {\bf 58} (2008) 51;
M. B. Voloshin, Prog. Part. Nucl. Phys. {\bf 61} (2008) 455;
N. Drenska, R. Faccini, F. Piccinini, A. Polosa, F. Renga and C. Sabelli, Riv. Nuovo Cim. {\bf 033} (2010) 633;
 N. Brambilla et al,  Eur. Phys. J. {\bf C71} (2011) 1534.


 \bibitem{WangHuangTao} Z. G. Wang and T. Huang,  Phys. Rev. {\bf D89} (2014) 054019.



\bibitem{One-gluon} A. De Rujula, H. Georgi and S. L. Glashow, Phys. Rev.  {\bf D12} (1975) 147;
 T. DeGrand, R. L. Jaffe, K. Johnson and J. E. Kiskis, Phys.  Rev.  {\bf D12} (1975) 2060.


\bibitem{Jaffe2003} R. L. Jaffe and  F. Wilczek, Phys. Rev. Lett. {\bf 91} (2003) 232003;  R. L. Jaffe, Phys. Rept. {\bf 409} (2005) 1.



\bibitem{WangDiquark} Z. G. Wang, Eur. Phys. J. {\bf C71} (2011) 1524;
  R. T. Kleiv, T. G. Steele and A. Zhang, Phys. Rev. {\bf D87} (2013) 125018.


\bibitem{SVZ79}  M. A. Shifman, A. I. Vainshtein and V. I. Zakharov, Nucl. Phys. {\bf B147} (1979) 385;
Nucl. Phys. {\bf B147} (1979) 448.

\bibitem{Reinders85} L. J. Reinders, H. Rubinstein and S. Yazaki, Phys. Rept. {\bf 127} (1985) 1.

\bibitem{WangHcHb} Z. G. Wang, Eur. Phys. J. {\bf C73} (2013) 2533.


\bibitem{Ioffe2005} P. Colangelo and A. Khodjamirian, hep-ph/0010175;
B. L. Ioffe, Prog. Part. Nucl. Phys. {\bf 56} (2006) 232.

\bibitem{PDG}  J. Beringer et al, Phys. Rev. {\bf D86} (2012) 010001.


\bibitem{Wang-NPA} Z. G. Wang, Nucl. Phys. {\bf A791} (2007) 106.

\bibitem{Ni-Da}  M. Nielsen, F. S. Navarra and S. H. Lee, Phys. Rept. {\bf 497} (2010) 4.

\bibitem{Wang4140} Z. G. Wang, Eur. Phys. J. {\bf C63} (2009) 115;
Z. G. Wang, Z. C. Liu and X. H. Zhang, Eur. Phys. J. {\bf C64} (2009) 373;
Z. G. Wang, Phys. Lett. {\bf B690} (2010) 403;
Z. G. Wang and X. H. Zhang,  Commun. Theor. Phys. {\bf 54} (2010) 323;
Z. G. Wang and X. H. Zhang, Eur. Phys. J. {\bf C66} (2010) 419.

\bibitem{Maiani-3872} L. Maiani, F. Piccinini, A. D. Polosa and V. Riquer, Phys. Rev. {\bf D71} (2005) 014028.

\bibitem{Ebert-3872} D. Ebert, R. N. Faustov and V. O. Galkin, Phys. Lett. {\bf B634} (2006) 214.


\bibitem{Carlucci-J0} M. V. Carlucci, F. Giannuzzi, G. Nardulli, M. Pellicoro and S. Stramaglia, Eur. Phys. J. {\bf C57} (2008) 569.

\end{thebibliography}
\end{document}